\documentclass[useAMS]{mn2e}
\usepackage{amsmath}
\usepackage{url}
\usepackage{amsfonts}
\usepackage{amsbsy}
\usepackage{subfigure}
\usepackage{verbatim}
\usepackage{amssymb}
\usepackage{amsbsy}
\usepackage{graphicx}
\usepackage{array}
\usepackage{breqn}
\usepackage{color}

\usepackage[T1]{fontenc}
\usepackage{ae,aecompl}

\usepackage{tabularx,ragged2e,booktabs,caption}
\newcolumntype{C}[1]{>{\Centering}m{#1}}

\title{MRI turbulence and thermal instability in accretion disks}
\author[Ross, Latter \& Tehranchi]{Johnathan Ross$^{1}$\thanks{E-mail:
   jpjr2@cam.ac.uk},
   Henrik N. Latter$^{1}$ and Michael Tehranchi$^{2}$\\
$^{1}$ DAMTP, University of Cambridge, CMS, Wilberforce Road,
Cambridge CB3 0WA, UK\\
$^{2}$  Statistical Laboratory, DPMMS , University of Cambridge, CMS, Wilberforce Road,
Cambridge CB3 0WA, UK\\}
\date{}

\begin{document}

\maketitle

\begin{abstract}
A long-standing puzzle in the study of black-hole accretion concerns the
presence or not of thermal instability. Classical theory
predicts the
encircling accretion disk is unstable, as do
self-consistent MHD simulations of the flow.
Yet observations of strongly accreting sources generally fail to exhibit
cyclic or unstable dynamics on the expected timescales.
This paper checks whether turbulent fluctuations impede
thermal instability. It also asks if it makes sense to conduct
linear stability analyses on a turbulent background. 
These issues are explored with a set of MRI simulations in thermally
unstable local boxes
in combination with stochastic equations that approximate the
disk energetics. These models show that the disk's thermal behaviour deviates
significantly from laminar theory, though ultimately a thermal runaway
does occur. We find that the disk temperature evolves as a biased random
walk, rather than increasing exponentially, and thus generates 
a broad spread of outcomes, with instability often
delayed for several thermal timescales. We construct a 
statistical theory that describes some of this behaviour, emphasising
the importance of the `escape time' and its associated probability distribution.
In conclusion, turbulent fluctuations on there own cannot stabilise
a disk, but they can weaken and delay thermal instability. 
\end{abstract}


\section{Introduction}

The assumption that the turbulent stress is proportional to
pressure in accretion disks (the $\alpha$-model) is a 
fundamental ingredient of classic
accretion disk theory (Shakura \& Sunyaev
1973). It has been partially justified by
successful application to quasi-steady systems, such as the thermal
spectrum of dwarf novae and X-ray binaries 
(e.g.\ Warner 1995, Gierli\'{n}ski \& Done 2004);
but when
applied to more delicate time-dependent dynamics,
such as instabilities, the model has encountered difficulties. 
For instance, the modelling of dwarf novae outbursts requires
different alphas for the high and low states (Smak 1984).
Another example involves
radiation-pressure dominated accretion flows which the alpha model predicts 
are subject to
thermal and viscous instability
(Shakura \&
Sunyaev 1976, Lightman \& Eardley 1974). X-ray observations, however, fail to find
variability on the timescales expected 
(Gierli\'{n}ski \& Done 2004), with only the exceptional luminous source
GRS 1915+105 and the intermediate black hole HLX-1 exhibiting
anything like cyclic behaviour driven by thermal instability (Belloni
et al.~1997, Done
et al.~2004, Sun et al.~2016, Wu et al.~2016). While it is possible
the disks are stabilised by an additional but unknown
 cooling mechanism, it may be that the heating
depends on temperature in a weaker way than the alpha model
assumes. For example, the turbulent stresses may be proportional to gas
pressure rather than total pressure, or not on pressure at all
(Gierlinski \& Done 2004). 

Of course, one can bypass the alpha model (and its assumptions) and 
simulate the MHD turbulence in these disks self-consistently, the
turbulence then supplied directly by the magnetorotational instability (MRI, Balbus
\& Hawley 1991). 
And in fact early work indicated that radiation-pressure dominated flows were
thermally stable (Hirose et al.~2009), in agreement with most
observations.
However, recent local and global simulations \emph{do} exhibit thermal runaways
(Jiang et al.~2013, Sadowski 2016, Mishra et al.~2016), 
though these expose additional complications that may
suppress instability, such as
numerical effects (especially box size) and the impact of a mean
magnetic flux. 
Obviously, both observations and simulations indicate that the onset and
development of thermal instability is far less straightforward than
predicted by the classical laminar theory. 

One very clear complication is the fact that the
stress-pressure relationship can change,
for both numerical and physical reasons, yielding
instability or stability depending on conditions in the disk and
in the simulation. 
A quite separate issue is the assumption that thermal instability can
be appropriately defined at all, at least when dealing with quasi-steady
turbulent states. The alpha-theory treats the turbulence as a static
eddy viscosity, and hence the equilibrium state as laminar. However, if the state hosts
vigorous fluctuations it may not be well-defined, or  even make sense,
to add a small linear
perturbation on top the stochastic background field
and subsequently
calculate
a growth rate.
One
envisages that, at the very least, non-model, non-exponential behaviour
ensues. Indeed, Jiang et al.~(2013) report delayed runaway and algebraic
growth rather than exponential growth in their MHD simulations, while
Janiuk \& Mishra (2012) show via a stochastic 1D model
that fluctuations induce random luminosity variations rather than the
regular outbursts expected.
It is to this aspect of the problem that this paper is devoted,
focussing
on the constructive and destructive interference of turbulence
on thermal instability.

In order to isolate the essence of the problem
we employ an idealised model of MRI turbulence and of
thermal instability. Unstratified shearing box simulations are performed
using the code RAMSES on a state that is MRI turbulent and thermally
unstable (at least according to the laminar alpha theory). Note that radiation pressure
is omitted and the gas cools due to a simple cooling function. 
We find that the turbulent fluctuations induce
thermal behaviour substantially different to that expected from
laminar theory.
In
particular, the evolution of the temperature resembles more a biased random
walk than an exponential runaway, with
a wide range of trajectories possible: the temperature in some simulations
departs from the laminar equilibrium relatively rapidly,
whereas in others it can `hang around' for several thermal times. 

This motivates a probabilistic interpretation of instability, and we
develop a simple statistical framework based on
the model of geometric Brownian
motion.
A key idea is that of the `escape time' $t_{\text{esc}}$ (which replaces the e-folding
time). It describes how long it takes for the system to deviate
significantly from the equilibrium. Reduced models involving both
white noise and the power spectrum of the MRI show that the
probability distribution of $t_{\text{esc}}$ possesses a long tail.
Thus there is a reasonable chance in any given
simulation that thermal runaway is delayed.
It should be stressed that ultimately realistic models of
disks still undergo
thermal runaways: turbulent fluctuations can impede instability but it
cannot destroy it. The stabilisation witnessed in Janiuk \& Misra
(2012) we attribute to the peculiarities of their stochastic model and
a very large noise amplitude.

Another feature of our MHD simulations is thermal fragmentation 
when the cooling rate is too small,
and hence the laminar thermal instability timescale too short. The
disk can then break up into hot and cold clouds. This occurs when the
thermal mixing (by turbulence or radiative diffusion) is inefficient
compared to thermal instability. On a sufficiently long lengthscale
this is always presumably the case, but how this relates to the onset
of instability in hot
accretion flows is unclear. Estimates of both radiative diffusion and turbulent mixing suitable 
for the inner regions of X-ray binaries indicate that fragmentation is
a marginal possibility.

The paper is organised as follows. In Section 2 we discuss a number of
issues pertinent to thermal instability, stochastic
fluctuations, and the limits of the alpha theory (not all of which we
take up in the paper). The third and fourth sections contain our numerical MHD
model and the corresponding results, respectively. We then explore
stochastic models in Section 5 and construct a statistical theory to
help explain the MHD
simulations. Our conclusions are then presented in Section 6.

\section{Theoretical issues}

\subsection{Stress-pressure relationship}

The exact dependence of the turbulent stress, $\Pi_{xy}$, on pressure 
is key to the onset of thermal stability in radiation pressure dominated flows. 
The instability occurs when $\Pi_{xy}\propto
(P_{\text{gas}}+P_{\text{rad}})$, but does not occur
when $\Pi_{xy}\propto P_{\text{gas}}$ (Piran
1978). Here $P_{\text{gas}}$ and $P_{\text{rad}}$ denote gas and
radiation pressure, respectively. The failure to observe
signatures of thermal instability in most X-ray observations have been
attributed to the stress depending on gas pressure alone, or possibly
the geometric mean of gas and radiation pressure (Gierlinski \& Done
2004).
It has
also been speculated that only exceptionally luminous flows could lead to a
situation where $\Pi_{xy}\propto P_{\text{rad}}$, explaining the
outbursts of GRS 1915+105 (though what exactly causes this shift in the
stress's behaviour is unclear) (Gierlinski \& Done 2004, but see also 
King and
Ritter 1998). 
Numerical simulations of the MRI have since
complicated this picture, as they indeed exhibit instability 
(Jiang et al.~2013),
and we are left with the task of numerically tracing out 
the non-straightforward behaviour of $\Pi_{xy}$ in different conditions.

While earlier unstratified shearing box simulations without radiation
pressure found only a weak dependence of stress on pressure (Sano et
al. 2004), recent work has shown that $\Pi_{xy}\propto P_{\text{gas}}$
when the following conditions are satisfied: the computational
domain is sufficiently large, explicit dissipation is included, and
there is no net magnetic field (Ross et al.~2016, hereafter RLG16). 
Small boxes restrict the size of the turbulent eddies and
prevent them from fully responding to an increase in
pressure (be it gas or radiation). This restriction no doubt played a role
in the failure of early radiation-pressure dominated simulations to
show thermal instability (Hirose et al.~2009): the radial
domain was too small, the eddies unnaturally
confined, and as a result the stress unable to respond to changes in
the total pressure.
When larger radial boxes are used, as in recent work, the instability
does in fact materialise (Jiang et al.~2013).

Another numerical effect uncovered by RLG16 was a sensitivity to the
grid in simulations with no net magnetic field and
no explicit diffusion. As in isothermal runs
(Fromang \& Papaloizou 2007), the stress is proportional to grid
size and this leads to a significantly weaker stress-pressure
relationship. 
Local boxes are not the only domains that exhibit the effect;
recent vertically stratified simulations also show that
the stress depends on the grid length (Ryan et al.~2017). 
The weaker dependence, both from numerical dissipation and from the
box size, leads to artificially more stable systems (as can be shown
from dimensional analysis). It
is therefore necessary for both of these numerical complications 
to be considered when simulating thermal instability involving gas pressure. 
It is likely that global disk simulations of the
MRI also suffer from strong numerical effects, though these have
yet to be fully explored.

Another intriguing result from RLG16 is that the stress-pressure 
relationship depends on the existence and strength of any imposed magnetic flux.
The stronger a mean toroidal flux, the weaker the relationship. If this
behaviour generalises to other field configurations and to
the radiation-dominated regime, then one
might speculate that highly magnetised flows are less prone to thermal
instability. Indeed global simulations of the MRI
suggest that strong magnetic fields impede thermal instability (Sadowski 2016), 
possibly because they weaken the connection between the stress and
pressure. This raises the interesting prospect that to assess
susceptibility to thermal instability we must also account for the
build up and evacuation of large-scale magnetic flux, in addition to
the turbulent dynamics (e.g.\ Guilet \& Ogilvie 2012a, 2012b). 

\subsection{Time lags}

The alpha theory is a turbulence closure model,
supplying a simple eddy viscosity in place of the complicated
and chaotic time-dependent dynamics of the flow. On timescales and
lengthscales much longer than the characteristic scales of the turbulence,
this approximation provides an adequate description, but its
performance worsens the shorter the scales of interest. Since thermal
instability can possess growth rates of tens of orbits (or less), then the
detailed turbulent dynamics could potentially interfere with its onset.

Simulations, as expected, show that on shorter timescales the stress-pressure
relationship is more complicated than a mere proportionality.
One interesting feature is a time lag of a few orbits 
between the stress and
pressure. Moreover, it is the pressure that follows the stress on
shorter times,
rather than the other way around. And so the dependency is opposite 
to that assumed by the alpha theory:
bursts in stress are followed by jumps in pressure (Hirose et
al.~2009). 
What is happening here is that
the bursts in stress drive fluctuations in the heating rate 
(once their associated energy has reached the dissipation scale)
and hence cause bursts in pressure a short time later.

The effect of this time delay on a stable thermal
equilibrium was first explored by Hirose et al.~(2009), who argued
that if the direction of causation was from the stress to the pressure
then thermal instability may not work. The argument fails, however,
to acknowledge that the stress-pressure dynamics exhibit
different timescales, with longer timescales ($>10$ orbits) 
characterised by
a pressure-dependent stress (as in the classical theory), while the
short timescales show the lags described above (Latter \&
Papaloizou 2012, RLG16).

Follow up work by Lin et al.~(2011) and Ciesielki et al.~(2012)
present linear instability analyses of an alpha disk model with a
time-delayed alpha. These show that the simulated delay of 1-10 orbits is
insufficiently long to stabilise a thermally unstable state. They also
reveal potential inconsistencies in such simple models: for example,
infinite growth rates are possible for certain time-lags,
a physical impossibility. Of course,
these models are also incomplete, as they include only the 
either the short term or long term dependencies, but not both concurrently.

\subsection{Temporal  fluctuations}

MRI turbulence has strong variation over a range of timescales, from
tens of orbits 
to a few
shear times (Sano
et al.~2004, Lesur \& Ogilvie 2008). In what sense does a thermal equilibrium exist in such a
system? Time varying perturbations are constantly emerging which
lead to a shifting balance of heating and cooling
that the system continually responds
to. On the other hand, if we assume that there is well-defined mean
equilibrium, 
then it is
awash in finite amplitude fluctuations. How can one then undertake a linear
instability analysis? Is it meaningful to add a tiny perturbation
ontop of a sea of 
finite amplitude perturbations and check if it grows or not? 
On long length and time scales this might work, but
certainly not on shorter scales.

Putting aside the difficulty of interpreting linear stability
analyses, a stochastic system exhibits a range of complicated and
sometimes
unexpected behaviour. A classic example is the destabilisation of
fixed points deemed stable by laminar theory. Originally studied in
biological population dynamics (e.g.\ Levins 1969, May 1973), this
feature of noisy systems appears in numerous
applications, such as atmospheric modelling (e.g.\ De Swart \& Grasman
1987), where the unresolved short time and lengths dynamics are
represented by stochastic terms (see Majda et al.\ 1999, 2003).
On the other hand, 
the influence of stochasticity on an otherwise \emph{unstable} fixed point
has been studied in financial mathematics, where geometric Brownian motion can
be used to model volatile stock prices in a rising market. Despite the
mean trend of increasing prices, stochasticity can depress the price
of some stock dramatically, if not stabilising the fixed point then
delaying a runaway in price for some period of time.

The very last example is perhaps the most relevant for our study of
thermal instability, as it possesses the key ingredients of (a) an
unstable fixed point (according to a deterministic or `laminar'
theory) and (b) stochastic fluctuations. The competition between them
gives rise to behaviour one might
liken to a biased random walk. In between the kicks delivered by the
turbulence, the system drifts according to the deterministic unstable
dynamics. One can then imagine certain limits: when the characteristic
frequency of the turbulence is much greater than the thermal
instability growth rate then we may expect an unbiased random walk,
and the system will only weakly sense the underlying thermal physics.
In the opposite limit, when the growth rate is much greater than the
turbulent frequency, the deterministic laminar
dynamics should be reproduced. It is in the intermediate regime, explored in this paper,
that interesting nontrivial behaviour manifests. There are
also other key ratios, such as the size of the kicks relative to the
magnitude of the fixed point or the initial condition. 
If these are too small, then we
return to the laminar case. But for intermediate values, as exhibited
by our MRI simulations, system trajectories can deviate markedly from
both the laminar behaviour and a simple unbiased random walk.   

\subsection{Spatial  fluctuations}

In the previous subsection we considered only temporal fluctuations on
the system variables, implicitly regarding them as `box averaged' or
mean quantities. Indeed, the $\alpha$ model assumes a homogenised
temperature over $\sim H$, the disk scale height.
Turbulent heating, however, is spatially inhomogeneous with strong
dissipation occurring in current and vorticity sheets and minimal
dissipation in the surrounding regions. 
These spatial fluctuations
also complicate the picture of thermal instability, especially when
the instability growth rate is large.

Some form of thermal mixing is
necessary to homogenise the temperature of the fluid. This can take
multiple forms, such as turbulent advection, radiative diffusion, or
thermal conduction. When thermal instability is present, the assumption of a
uniform temperature is reasonable as long as the instability time scale is
longer than the mixing timescale, $t_{\text{inst}}>t_{\text{mix}}$. For less
efficient mixing or stronger instability, there exists a maximal thermal coupling
length scale, $l\sim v_{\text{mix}}t_{\text{inst}}$. Regions separated by more than
this only weakly interact thermally during an efolding time. This implies
that regions of a disk separated by more than $l$ can undergo thermal runaways
independently, and the disk fragments into cold and hot clouds. 
 Regions of strong kinetic and magnetic dissipation are
likely to heat catastrophically, while those with weak dissipation will
cool catastrophically. How relevant this
scenario is in realistic disks is unclear, though perhaps marginally
possible in X-ray binaries. It is certainly possible in
numerical simulations as we show later.

\begin{figure}
\includegraphics[width=8cm]{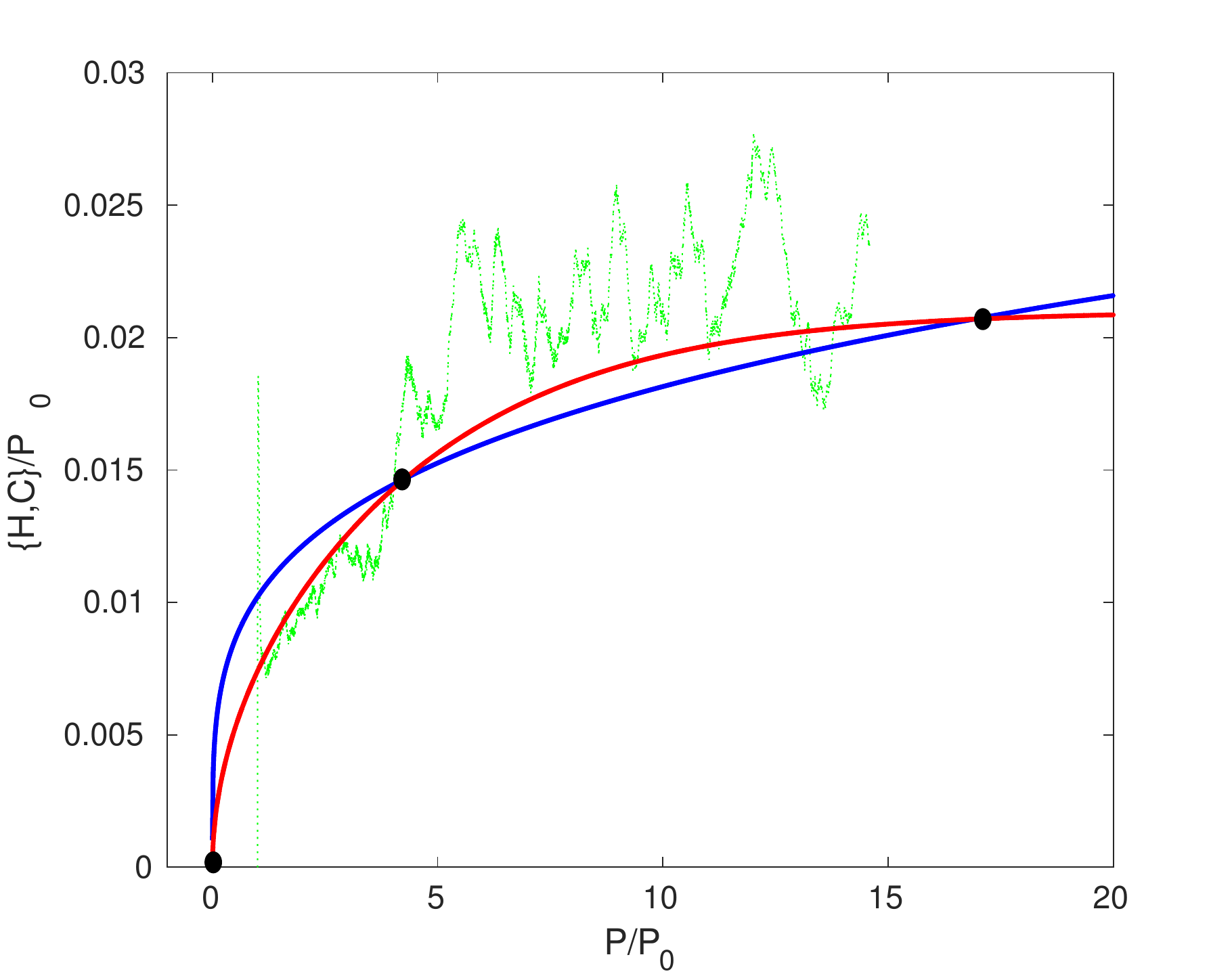}
\caption{For illustrative purposes, the red curve shows the fitted
  heating rate calculated from a zero net-flux simulation with
  resolution $\frac{1}{64}H_{0}=\Delta$ from RLG16 (shown in green).  
  The blue line is an example cooling function $\propto P^{0.25}$. 
  The fixed points are shown by black dotes.}
\label{Fig:HandC}
\end{figure}

\section{Numerical tools and setup} %3

\subsection{Formulation}

We wish to explore the essential features of thermal stability and its
onset in MRI driven turbulence and so we choose an idealised set-up to
isolate it. We adopt the local shearing box model (Goldreich \&
Lynden-Bell 1965). To prevent complications such as buoyancy, mass
loss and disk expansion we consider the unstratified case. With this
model, MRI driven turbulence can be obtained when a Keplerian flow
profile is assumed (Hawley et al.~1995). As is conventional, $x,y,z$
are the radial, azimuthal and vertical spatial variables and
$\boldsymbol{\hat{e}}_{x}$, $\boldsymbol{\hat{e}}_{y}$,
$\boldsymbol{\hat{e}}_{z}$ are the corresponding unit vectors. This
frame of reference co-rotates with the disk at some radius
with angular frequency
$\mathbf{\Omega}=\Omega\boldsymbol{\hat{e}}_z$.
The ideal compressible MHD equations are hence
\begin{align}
\label{eq::1}
&\frac{\partial \rho}{ \partial t} + \nabla \cdot ( \rho \boldsymbol{v}
) = 0, \\
&\rho \frac{\partial \boldsymbol{v}}{\partial t} +\rho (\boldsymbol{v}
\cdot \nabla)\boldsymbol{v} 
 = - 2 \rho \mathbf{\Omega} \times \boldsymbol{v}  +
3x\rho \Omega^{2}\boldsymbol{\hat{e}}_{x}-\nabla P \notag\\
& \hskip4cm  +( \nabla \times \boldsymbol{B} ) \times \boldsymbol{B} \\
&\frac{\partial \boldsymbol{B}}{\partial t} = \nabla \times
(\boldsymbol{v} \times \boldsymbol{B}), \\
&\frac{\partial \varepsilon}{\partial t} + \boldsymbol{v}\cdot\nabla
\varepsilon= -P\nabla\cdot\boldsymbol{v} +Q -\Lambda.
\label{eq::4}
\end{align}
where $\rho$ is the mass density, $\boldsymbol{v}$ is the velocity, 
 $P$ is the gas pressure, $\boldsymbol{B}$ is the magnetic field and the internal energy is
denoted by $\varepsilon$. Heating is represented by $Q$ and cooling by
$\Lambda$. 
This set of equations is then closed by relating the internal energy to the pressure by assuming an ideal gas so that
\begin{equation}
\varepsilon= P/(\gamma-1),
\end{equation}
where $\gamma$ is the adiabatic index, taken to be $7/5$. The sound speed is then given by $c_{s}=(\gamma P/\rho)^{1/2}$ and the
pressure scale height by $H=(2/\gamma)^{1/2}c_{s}/\Omega$.

Ideally we would include viscosity and Ohmic diffusion and so $Q$ would
be given by the sum of the physical dissipative processes; 
however, to resolve
the diffusion length scales requires a higher resolution than is
practical for this study. Instead, we rely on numerical dissipation
for heating. By solving Equations \eqref{eq::1} - \eqref{eq::4} in
conservative form, the kinetic and magnetic energy dissipated by the
grid is converted to internal energy. Energy that is extracted from
the background shear is converted to internal energy and ultimately
removed via the cooling function $\Lambda$. 

Finally, for our cooling function, we take a power law of pressure,
\begin{equation}
\Lambda = \theta\,P^{m},
\end{equation}
where $\theta$ and $m$ are both constants. Though this choice is
mainly for convenience, it might crudely approximate an optically thin medium.

\subsection{Numerical methods}

All of the simulations that we perform are carried out using RAMSES, a finite-volume Godunov code based on the MUSCL-Hancock algorithm 
(Teyssier 2002; Fromang et al.~2006). The HLLD Riemann solver 
(Miyoshi \& Kusano 2005),
 and the multidimensional slope limiter described in Suresh (2000) are used in all the simulations presented in this paper.

Rather than solving for the total $y$-momentum, we evolve the equivalent
conservation law for the angular momentum fluctuation $\rho
v_y^{\prime} =\rho( v_y - v_K)$,
 with $v_K$ the Keplerian velocity. An upwind solver is used for solving the azimuthal advection arising from $v_{K}$.
 The tidal and Coriolis forces are treated as source terms and
implemented following the Crank-Nicholson algorithm described in
Stone \& Gardiner (2010). 

 The algorithm solves for the total fluctuation energy $E^{\prime} \equiv \varepsilon + \rho
v^{\prime 2}/2 + B^2/2 $ and its conservation law is written as
\begin{align} \label{eq::7}
\frac{\partial E^\prime}{\partial t} +  \nabla\cdot &\left(E^\prime{
    \boldsymbol{v}^\prime} + {\boldsymbol{v}^\prime} \cdot \mathsf{P} \right) = -v_K\frac{\partial E^\prime}{\partial y} \notag\\
    &+ \left(B_xB_y - \rho v_x v_y^{\prime} \right)\frac{\partial v_K}{\partial x} -\Lambda,
\end{align}
where $\mathsf{P}$ is the total pressure tensor
\begin{align}
\mathsf{P} = (P + B^2/2)\sf{I} - \boldsymbol{B}\boldsymbol{B}.
\end{align}
The left hand side of Equation \eqref{eq::7} comprises the usual energy conservation law,
which we solve using the MUSCL-Hancock algorithm. 
The treatment of the two terms on the right hand
have been modified: the azimuthal 
advection of energy is solved with an upwind solver, and the second
term involving the Maxwell and 
Reynolds stresses is added as a source term.

For the set of simulations shown in this paper, we used a box size of
$(L_{x},L_{y},L_{z})=(4,5,4)H_{0}$  with a resolution of
$(N_{x},N_{y},N_{z})=(128,100,128)$. $H_{0}$ is a reference scale
height which is close to but not exactly the same as the scale height at the start of
a simulation. It will be defined in detail later. 
The grid scale is defined to be $\Delta=L_{x}/N_{x}$. We
set $\Omega=10^{-3}$ and $c_{s0}=10^{-3}$ in code units, where
$c_{s0}$ is the initial sound speed. The resolution is low in
comparison to other MHD shearing box simulations, however it is
sufficient to capture the basic properties of the MRI, in particular
the turbulent  fluctuations. Importantly, it is computationally
inexpensive
allowing for the simulations to run for $\gtrsim 500$ orbits.

\subsection{Thermal equilibrium}\label{Sec:TE}

In our shearing box model, the energy is injected into the
computational domain by the second term on the right side of Equation \eqref{eq::7} which
represents the liberation of shear energy by the total stress:
\begin{equation}
\Pi_{xy}=B_xB_y - \rho v_x v_y^{\prime}.
\end{equation}
Simulations have found that the box-averaged stress $\langle \Pi_{xy}
\rangle$ is roughly proportional to the 
box-averaged gas pressure to a given power (Sano et al.~2004, RLG16),
though be aware of the caveats given in Section 2. 
Therefore, in numerical simulations, the averaged
heating rate may be approximated by 
\begin{equation}
\langle Q\rangle\propto \langle\Pi_{xy}\rangle =\tilde{\alpha}\langle P\rangle^{q},
\label{eq::9}
\end{equation}
where $\tilde{\alpha}$ and $q$ may be calculated from the simulation.
Note that the former is
\textit{not} the same as the $\alpha$ parameter.
The exponent $q$ depends on the field geometry as well as the
numerical parameters (RLG16). 
For our set-up, $q=0.5$, as long as the pressure is sufficiently
small (see later).  

Combining this approximation with our cooling prescription,
the evolution of the volume averaged pressure, $\langle P\rangle$, is determined by
\begin{align}
\label{eqn::10}
\frac{d\langle P\rangle}{dt}&\approx\left(\gamma-1\right)\frac{3}{2}\Omega\langle\Pi_{r\phi}\rangle-(\gamma-1)\theta \langle P\rangle^{m} \\ 
&\cong a_1\langle P\rangle^{q}-a_2 \langle P\rangle^{m}.
\label{eqn::11}
\end{align}
where 
\begin{align}
a_1=\frac{3}{2}\Omega(\gamma-1)\tilde{\alpha}, \qquad a_2=(\gamma-1)\theta.
\end{align}
In the formulation of Equation \eqref{eqn::10} we have made the
approximation that $\langle P^{m}\rangle\cong\langle P\rangle^{m}$,
which is reasonable as long as the variance of $P$ within the box at
fixed time is not too large. This system has two fixed points 
\begin{align}
P_{1}&=0, \\
P_{2}&=\left(\frac{a_{1}}{a_{2}}\right)^{1/(m-q)}=\left(\frac{3\Omega\tilde{\alpha}}{3\theta}\right)^{1/(m-q)}.
\end{align} 

In shearing box simulations the thermodynamics have an additional
complication: stress is independent of pressure once the scale height
is larger than the radial and vertical box sizes (RLG16). The exponent
$q$ may then be viewed as a function of the disk temperature, equal to
$0.5$ for cool disks, while asymptoting to 0 as the gas heats and the
scale height $H$ equals the box size $L_z$. 
This behaviour introduces the possibility of
an additional stable fixed point if $m<\max\{q\}$. In Fig.
\ref{Fig:HandC} we plot (a) the heating rate as a function of pressure from a $L=4H_{0},
\Delta=1/64$ simulation (appearing in RLG16), (b) a smooth fit to
this curve, and (c)
overlay a $m=0.25$ power law cooling. Where the latter two curves
intersect
 give three thermal equilibria. These are the $P_1$ and $P_2$ fixed
 points, described above, in addition to a
third equilibrium, $P_3$, which arises from the finite size of the
box. 
Note that a similar fixed point
will also appear in vertically stratified simulations.

The above volume averaged analysis is deterministic, but, turbulent
flows are not. The stress fluctuates around its mean value as a result
of the formation and break up of coherent structures within the
flow. This means that $Q$ can no longer be expressed in as simple a
form as Equation \eqref{eq::9}. Instead, the stress is determined by the
sum of a deterministic term and a fluctuating term. By using Equation
\eqref{eqn::10} we can obtain an estimate of the equilibrium pressure
that the system feels at any given time

\begin{equation}
P_{\text{exp}}=\left(\frac{3\Omega\langle \Pi_{xy}\rangle}{2\theta}\right)^{1/m}.
\label{Eqn::pexp}
\end{equation}

\subsection{Thermal instability}\label{Sec::Instab}

According to a linear analysis of Equation \eqref{eq::4}, 
the thermal instability criterion is 
\begin{equation}
\frac{dQ}{dP}>\frac{d\Lambda}{dP}
\end{equation}
which, given our power law expressions for $Q$ and $\Lambda$, 
leads to the simple condition $$q>m$$ for $P_{2}$, and the opposite
stability for $P_{1}$ and $P_3$. 

We calculate the growth rate as
\begin{equation}
s_g=(q-m)(\gamma-1)\left(\frac{3\tilde{\alpha}\Omega}{2}\right)^{(1-m)/(q-m)}\theta^{(q-1)/(q-m)},
\label{Eqn::therm_lam}
\end{equation} 
which can be used for comparison with the MHD simulations. The
associated instability timescale we define to be $t_{\text{inst}} =1/s_g$. 

If the system was truly laminar the instability criterion would be
fully determined by $p$ and $q$ and the instability
timescale from \eqref{Eqn::therm_lam}. Fluctuating systems, however,
exhibit non-exponential and indeed non-monotonic behaviour which is
poorly approximated by the laminar model. In particular, $t_{\text{inst}}$ may
be an unsuitable measure for the instability timescale. In its place we introduce
the `escape time' $t_{\text{esc}}$ of a simulation, 
which is defined to be the last instance that the system lies within a
pressure interval containing the fixed point. 
Mathematically it may be defined via
\begin{equation}
t_{\text{esc}}=\max\left\lbrace t>0 : \vert P(t)-P_{\text{eqm}}\vert=\delta\right\rbrace,
\label{Eqn::Tesc}
\end{equation}
where $P_{\text{eqm}}$ is one of the three equilibria introduced
earlier, and $\delta$ is the interval size, outside of which we
consider the system to have unequivocally departed from the fixed
point. We are free to specify the size of $\delta$, and it might
reflect the particular problem of interest. For disk transitions
between states differing by many orders of magnitude in temperature
we might be generous with $\delta$, permitting it to be up to 10 times
the fixed point pressure. For smaller transitions, then $\delta$ must
be smaller.
Note that the  system may stochastically dip in and out of this interval, but
$t_{\text{esc}}$ will capture the time when the system finally leaves it forever.
If the disk is laminar, then 
\begin{align}
 t_{\text{esc}}=t_{\text{inst}}\ln\left(\frac{\delta}{|P_{\text{eqm}}-P_{\text{init}}|}\right) 
\end{align}
where $P_{\text{init}}$ is the
initial pressure of the system.

\subsection{Initial conditions}  

If we want a thermal equilibrium to be achievable within our box and
for the stress to depend appreciably on pressure, the initial
conditions must be chosen carefully. An initial state that is too hot
means the box size will unduly influence the stress and weaken
$Q$'s dependence on $P$. 
As a consequence, thermal instability will fail to occur.

Stress is only observed to be
a strong function of pressure ($\Pi_{xy}\propto P^{1/2}$) 
when $\Delta\ll H<L$  (Sano et al.~2004,
RLG16). Therefore, $P_{2}$ must be sufficiently low so that the stress is
increasing with pressure, but sufficiently high so that
the characteristic lengthscale of the turbulence is not on (or below)
the grid. 
That way we can be assured that the thermal instability is able to
appear in our numerical set-up. We choose
\begin{equation}
\frac{\sqrt{P_{2}}}{\sqrt{P_{\text{box}}}}=\frac{H_{2}}{L}\approx 0.5
\label{eqn::P0Peqa}
\end{equation}
where $P_{\text{box}}$ is the pressure at which the scale height equals the
box size and $H_{2}$ is the scale height when $\langle P\rangle
=P_{2}$. From this we can obtain the required value of $\theta$ for a given choice of $m$:
\begin{equation}
\theta=\frac{(3/2)\tilde{\alpha}}{(0.25P_{box})^{(m-q)}}
\end{equation}

To achieve a turbulent state we first initialize a zero net-flux
simulation with an initial field of $\boldsymbol{B}=B_{0}\sin(2\pi
x)\mathbf{\hat{e}}_z$ and no cooling. In code units
$B_{0}=\sqrt{2/\beta}$ which we set with $\beta=10^{3}$. Once a
turbulent state is reached we switch on cooling with $\theta=10$ and
$m=2$. This choice of $m$ is to obtain a $P_{2}$ which we expect to be
stable, as discussed in Section \ref{Sec::Instab}. These parameters
lead to $P_{2}\approx 3.2P_{0}$ where $P_{0}$ is the initial pressure,
having used that $P_{\text{box}}\approx 16P_{0}$. During this steady state we
calculate $\tilde{\alpha}\approx2.3\times10^{-5}$.  
This fully turbulent state in thermal equilibrium is used as our
initial condition. The parameters $m$ and $\theta$ are then changed as
appropriate.

\begin{figure}
\includegraphics[width=10cm]{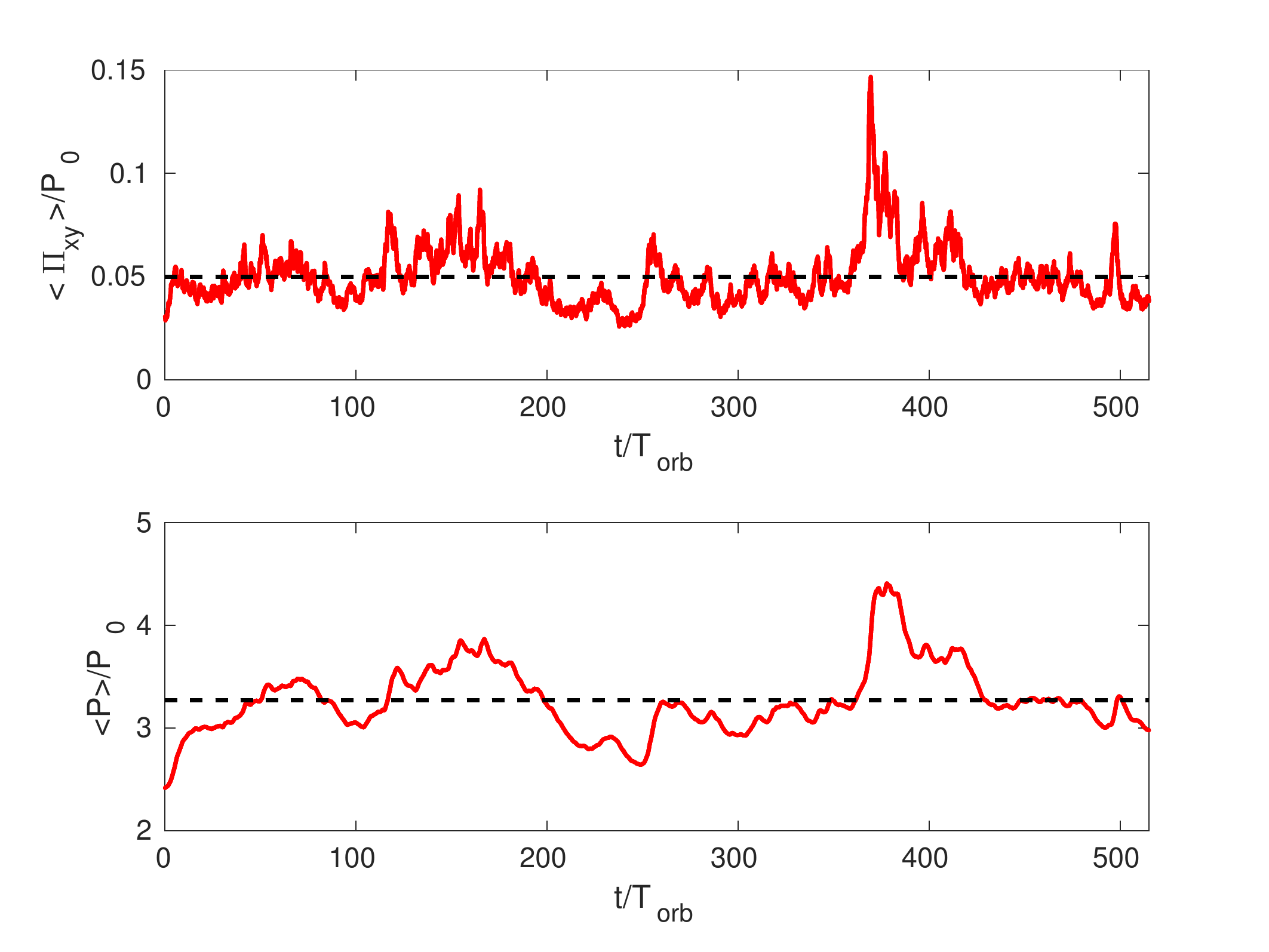}
\caption{The evolution of the box-averaged pressure and stress from simulation R1.}
\label{Fig::r1_p}
\end{figure}

%\begin{figure}
%\includegraphics[width=9cm]{images/Run1SvP-eps-converted-to.pdf}
%\caption{The evolution of the logarithm of the box averaged stress against the logarithum of the box averaged stress in R1.}
%\label{Fig::r1_svp}
%\end{figure}

\begin{figure}
\includegraphics[width=10cm]{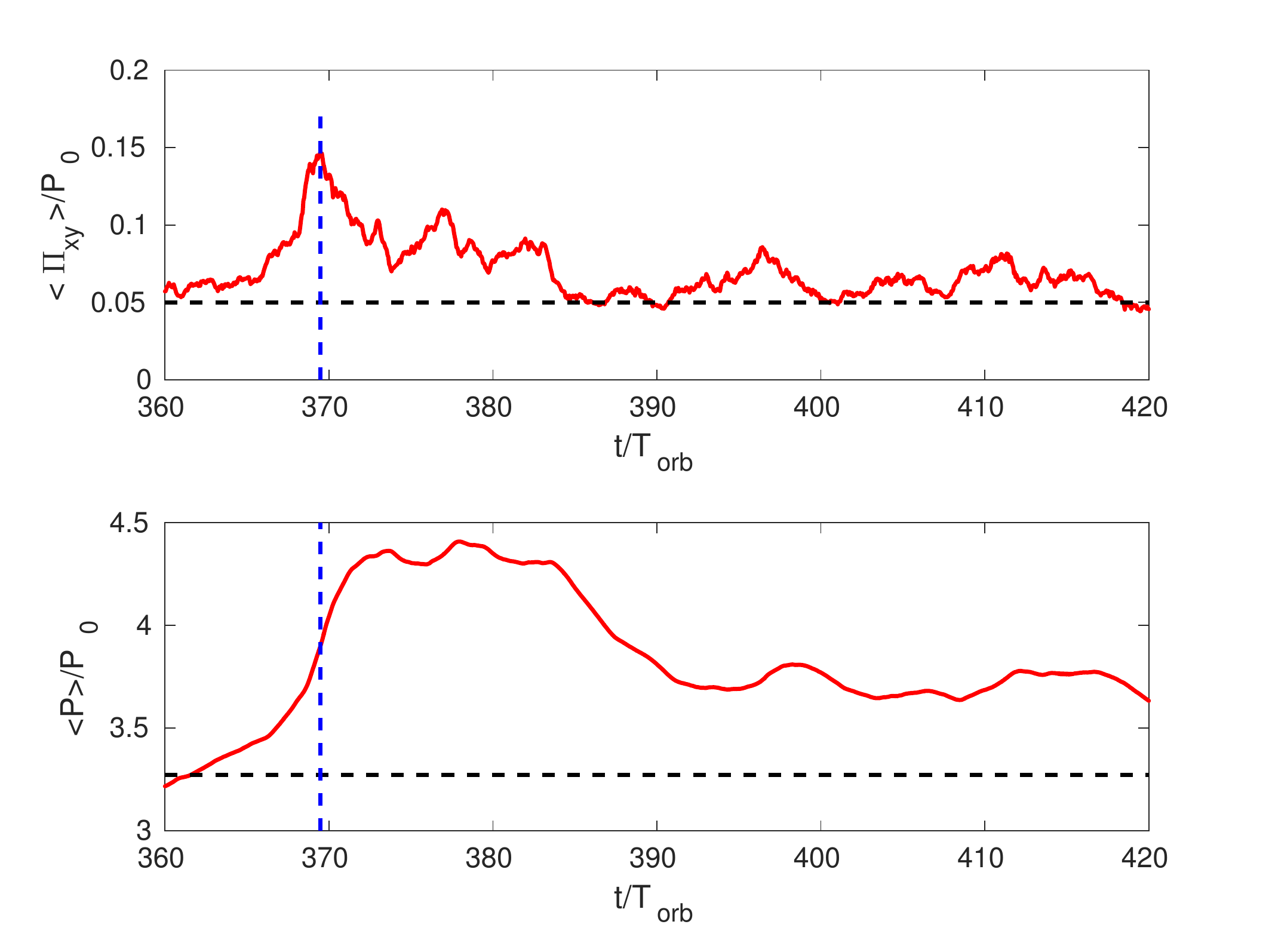}
\caption{A zoom-in of the pressure and stress evolution in
  Fig.~\ref{Fig::r1_p}. 
 The blue dashed vertical line indicates the time at which the stress
 achieves its maximum in this interval.}
\label{Fig::R1_P_Z}
\end{figure}

\section{Results}

\subsection{Stable system}

\begin{figure}
\includegraphics[width=8cm]{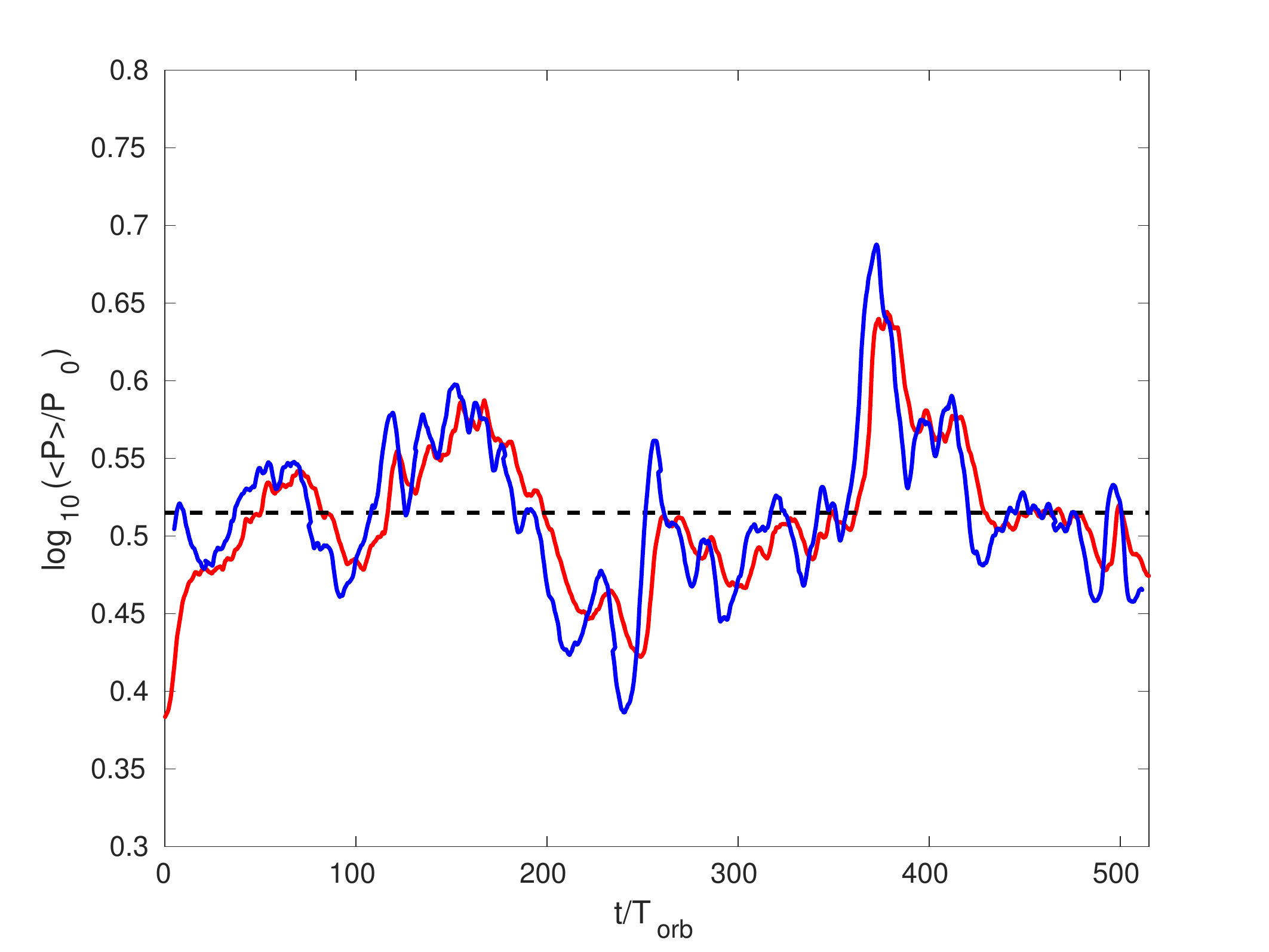}
\caption{The evolution of the logarithm of the box averaged pressure (red) along with the logarithm of the instantaneous expected equilibrium pressure calculated using an average stress over the previous $9t_{\text{orb}}$ (blue) for the thermally stable simulation R1. The black dashed line indicates the mean equilibrium pressure.  }
\label{fig::r1_pexp}
\end{figure}

To fix ideas we first consider the stable thermal equilibrium
associated with the initial condition, $\theta=10$, $m=2$ (simulation R1).
We run the simulation for over $500$ orbits and the resulting stresses
and pressures are shown in Fig. \ref{Fig::r1_p}. As predicted from
the laminar
linear analysis, the equilibrium is `stable' in the sense that both
the stress and pressure fluctuate within some interval enclosing 
$P_{2}$.
Though the stress shows substantial variation during the simulation (with a
maximum value of $\approx3$ times its mean), there is no runaway or
mean drift during the $500$ orbits. The variability in the pressure is
less extreme, with a maximum value of $\approx 1.5$ times its
mean value. 
 The turbulence administers
random `kicks' to the system, but the deterministic physics always 
draws it back to the vicinity of $P_2$. 
The stochastic fluctuations in stress are a result of
dynamo cycles and the formation and breakup of coherent
structures. The energy in the flow then cascades down to the
dissipation scale where the magnetic and kinetic energies are
converted to thermal energy leading to changes in  pressure.

A short time delay of a few orbits ($t_{\text{orb}}$) between the stress and pressure
is clearly visible in Fig. \ref{Fig::R1_P_Z}, a result of the finite
time taken for kinetic and magnetic energies to reach the dissipation
scale from the injection scales (Hirose et al.~2009). 
The variations in pressure are smoother and longer
than those in the stress. However, this short time dependency of pressure on
stress is distinct to
the longer time dependence of stress on pressure that drives
instability/stability (Latter \& Papaloizou 2012).

Though the pressure seems to be drawn back to the mean of $P$, 
in actual fact the equilibrium balance that the system feels at
any given instance is changing with time. A crude approximation to
this is the instantaneous fixed point $P_{\text{exp}}$,
calculated using Equation \eqref{Eqn::pexp}. But because the thermal
timescale is longer than the turbulent timescale, a better
approximation is the average of $P_{\text{exp}}$ over the last thermal time.
Using Equation
\eqref{Eqn::therm_lam} we estimate this time to be $\approx
9$ orbits. The resulting smoothed and time-dependent equlibrium is
plotted in Fig. \ref{fig::r1_pexp} in blue, alongside the actual
pressure of the system $\langle
P\rangle$ in red. 
It is clear that $\langle
P\rangle$ follows the equilibrium with a time lag of
$\lesssim 10$ orbits. Despite the fluctuations the
system senses the stable fixed point and is attracted towards it,
though because of those same fluctuations it can never come to rest
upon it.

\subsection{Unstable systems}

\begin{figure*}
\includegraphics[width=15cm]{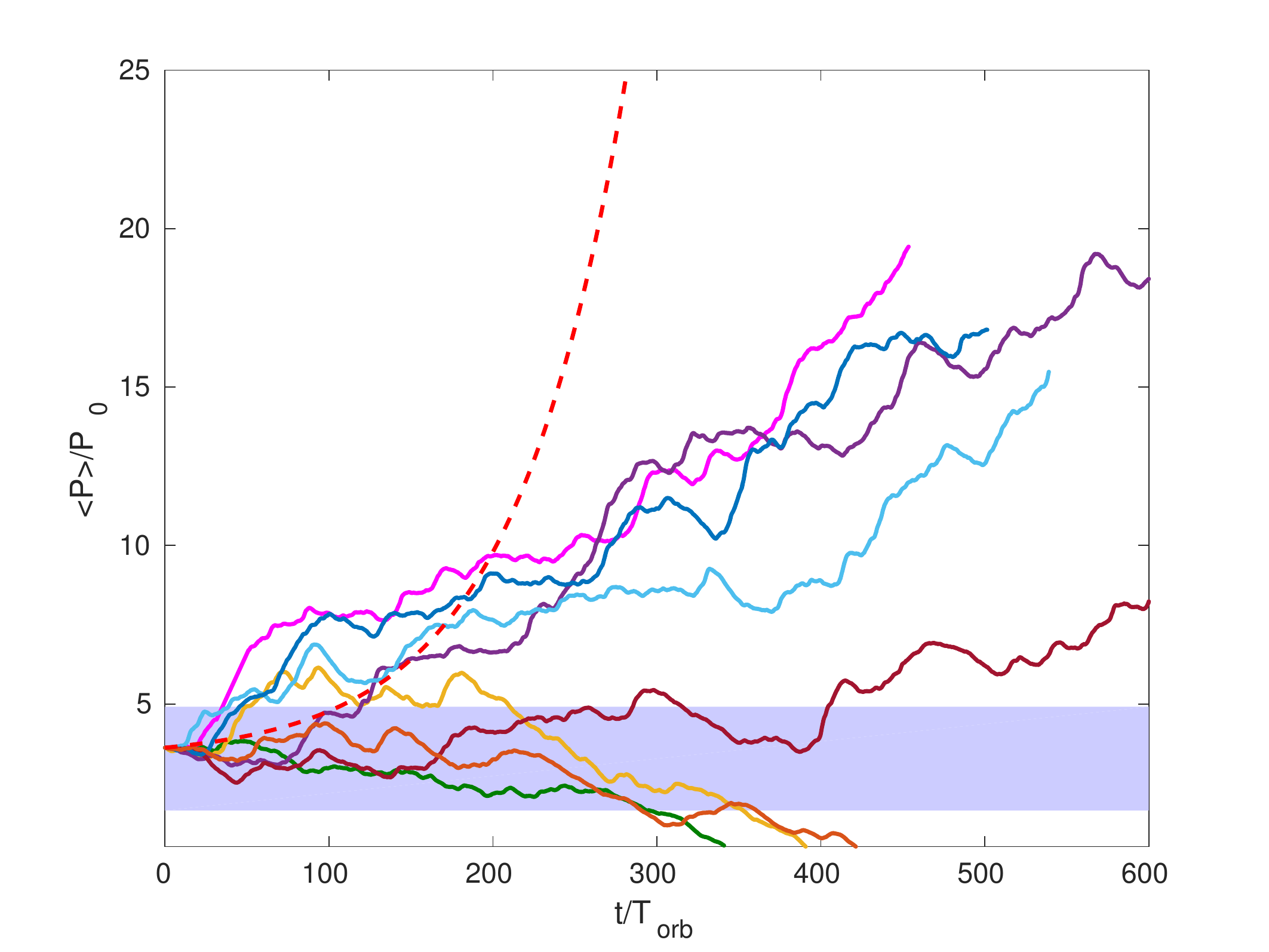}
\caption{Pressure verses time for the unstable simulations R2a-R2h shown by solid curves. The dashed red line is linear evolution. The filled area indicates the fixed point region $0.5P_{\text{eqm}}<P<1.5P_{\text{eqm}}$.}
\label{fig::UN_P_S}
\end{figure*}

Having explored the stable case we now proceed to an
unstable thermal equilibrium. For this set of simulations we choose
$m=0.25$ which we expect to be unstable 
based on the argument in Section \ref{Sec::Instab}. 

We present 8 simulations (R2a-R2h) that have the same fully turbulent
initial state. In order to vary the initial condition
between runs, these simulations each have slightly different
$\theta\approx1.31\times10^{-9}$ and so have slightly different
$P_{2}$. Each of these is within a few percent of the initial
pressure. In practice this means that though each simulation starts
from the same initial condition, each corresponds to a slightly
different perturbation from equilibrium.

In Fig. \ref{fig::UN_P_S} we show the evolution of the box-averaged
pressures alongside the trajectory derived from the laminar 
linear theory. Unsurprisingly the
turbulent fluctuations lead to a diversity of outcomes but do not
indefinitely prevent thermal runaway. For example, a large kick can
cause the system to escape from the fixed point on a shorter timescale
than the laminar timescale, or the system may remain close to the
fixed point for extended periods of time, longer than
$t_{\text{inst}}$.
Our most `stable' simulation remains close to equilibrium for
$400t_{\text{orb}}$.
 None
of the simulations can be well modelled by the laminar theory. For
$P\gtrsim5P_{0}$, the behaviour is closer to algebraic growth than
exponential runaway. In fact, the behaviour of the system is strongly
influenced by the fluctuations that occur on timescales comparable to
the instability timescale, $\sim 70 t_{\text{orb}}$ for these
parameters. 
A more apt description of
the system could be a biased random walk, when strong fluctuations in
stress repeatedly perturb the system, while in between kicks the
system drifts according  to the deterministic physics. 

If we consider some characteristic interval around the fixed point
 then we can find the very last time,
$t_{\text{esc}}$, the system was within this band, Equation
\eqref{Eqn::Tesc}. If we define the interval to be rather narrow,
$0.5P_{eq}<P<1.5P_{eq}$, then this `escape time' can be compared to
$t_{\text{inst}}$. We find a wide range of escape times in our MHD
simulations, from $0.25t_{\text{inst}}$ to $5t_{\text{inst}}$. Ideally, we would
calculate a probability distribution function for the escape time but
this would require substantially more simulations, which, at this time
is impractical. The choice of band is arbitrary, but  it should be chosen to suit the system.

Once $P>10P_{0}$ the simulations become cooler than
the laminar model prediction. This can be attributed to the box size
beginning to influence the evolution. At this point, the dependence of
stress on pressure decreases (see Fig. \ref{Fig:HandC}) and hence
the laminar model is an over estimate. This numerical effect
introduces a third equilibrium point, $P_{3}>P_{2}$, as discussed in
Section \ref{Sec:TE}. We do not observe a plateau in pressure
associated with the system being attracted to $P_{3}$, but, we expect
that if run for a sufficient duration then a  plateau would appear. 

That a runaway heating is limited by
the box size is an important problem, both for these simulations and
stratified radiation MHD simulations. To explore this further, we
rescale the simulations by choosing a larger value of $\theta$, and initialize a
simulation with the same initial turbulent state as in
R2a-R2h. However, now the initial condition is much further away from
the putative equilibrium, by a factor of some 3, and yet box effects
remain negligible, because 
Equation \eqref{eqn::P0Peqa} is still satisfied. In these runs
the thermal runaway was somewhat faster than witnessed in Fig 5 for
the hotter systems. These few simulations illustrate the numerical 
limitations inherent in any simulation of catastrophic heating
undertaken in a finite domain.

\begin{figure}
\includegraphics[width=8cm]{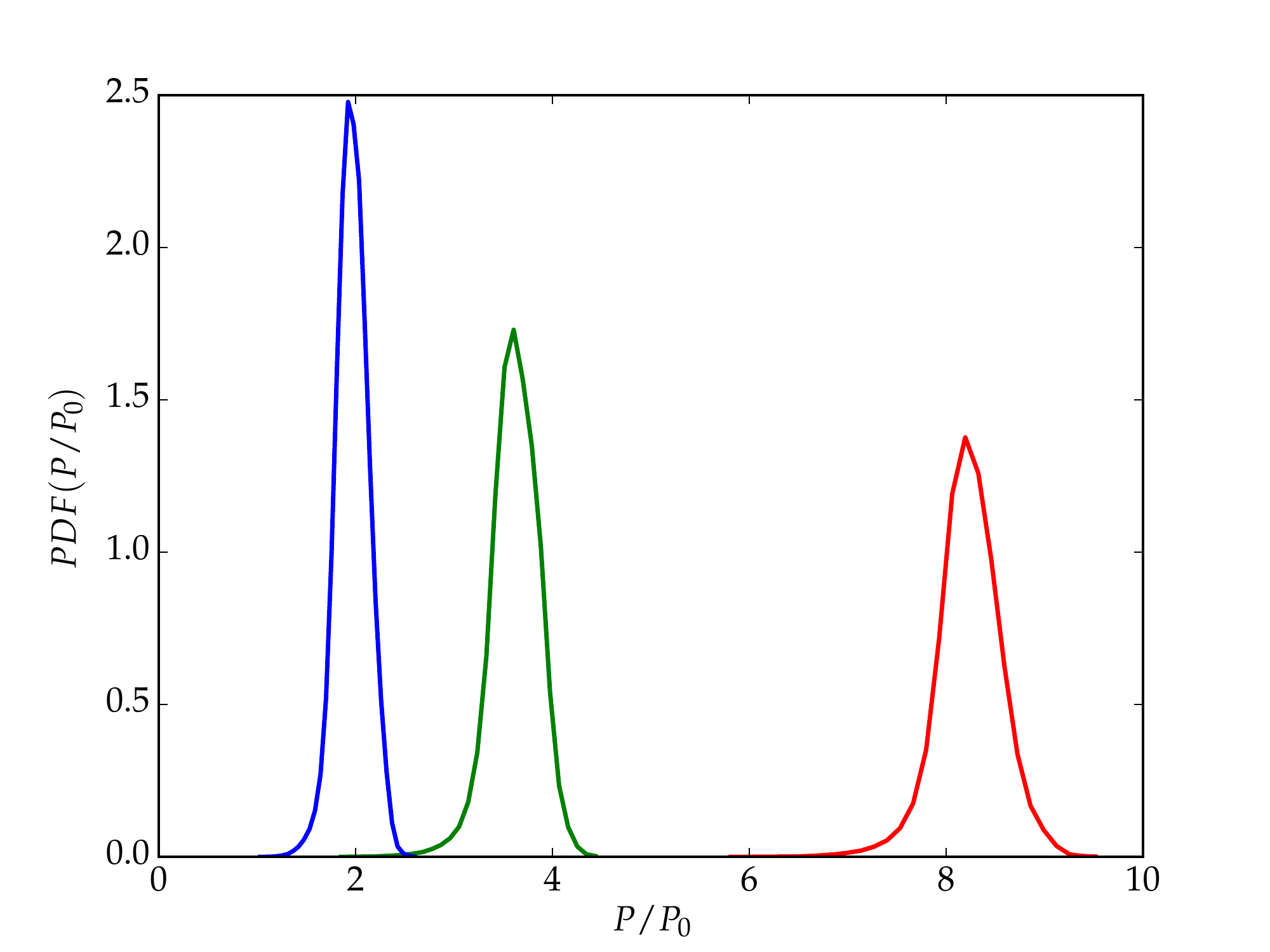}
\caption{The pressure probability distribution function $PDF(P/P_{0})$ from a heating (red) and a cooling run (blue) at $t=340t_{\text{orb}}$. The green curve shows the initial distribution.}
\label{fig::UN_pdf_P}
\end{figure}

\subsection{Thermal fragmentation}

In a turbulent system, heat is not deposited uniformly throughout the
box, but
rather is localised in coherent structures such as current and
vorticity sheets. If the instability timescale, $t_{\text{inst}}$, is large
compared to the mixing timescale, $t_{\text{mix}}$, then the
heating inhomogeneity will have little effect as temperature
fluctuations will be smoothed out. Conversely, if $t_{\text{mix}}>t_{\text{inst}}$
then regions of fluid can evolve independently of each resulting
in localised runaway. In Fig. \ref{fig::UN_pdf_P} we plot the
pressure probability distribution function, $PDF(P)$, for two R2
simulations at $t=340t_{\text{orb}}$. Inefficient turbulent mixing would
result in the spreading of $PDF(P)$ during runaway. The figure exhibits
minimal evidence of spreading in the heating run, and the pressure ensemble
evolves with a well defined and relatively narrow shape. The pressure in the  
cooling simulation is attracted to $P_{1}=0$ and in fact
 we see a further 
narrowing of the distribution as the equilibrium is approached.

\begin{figure}
\includegraphics[width=8cm]{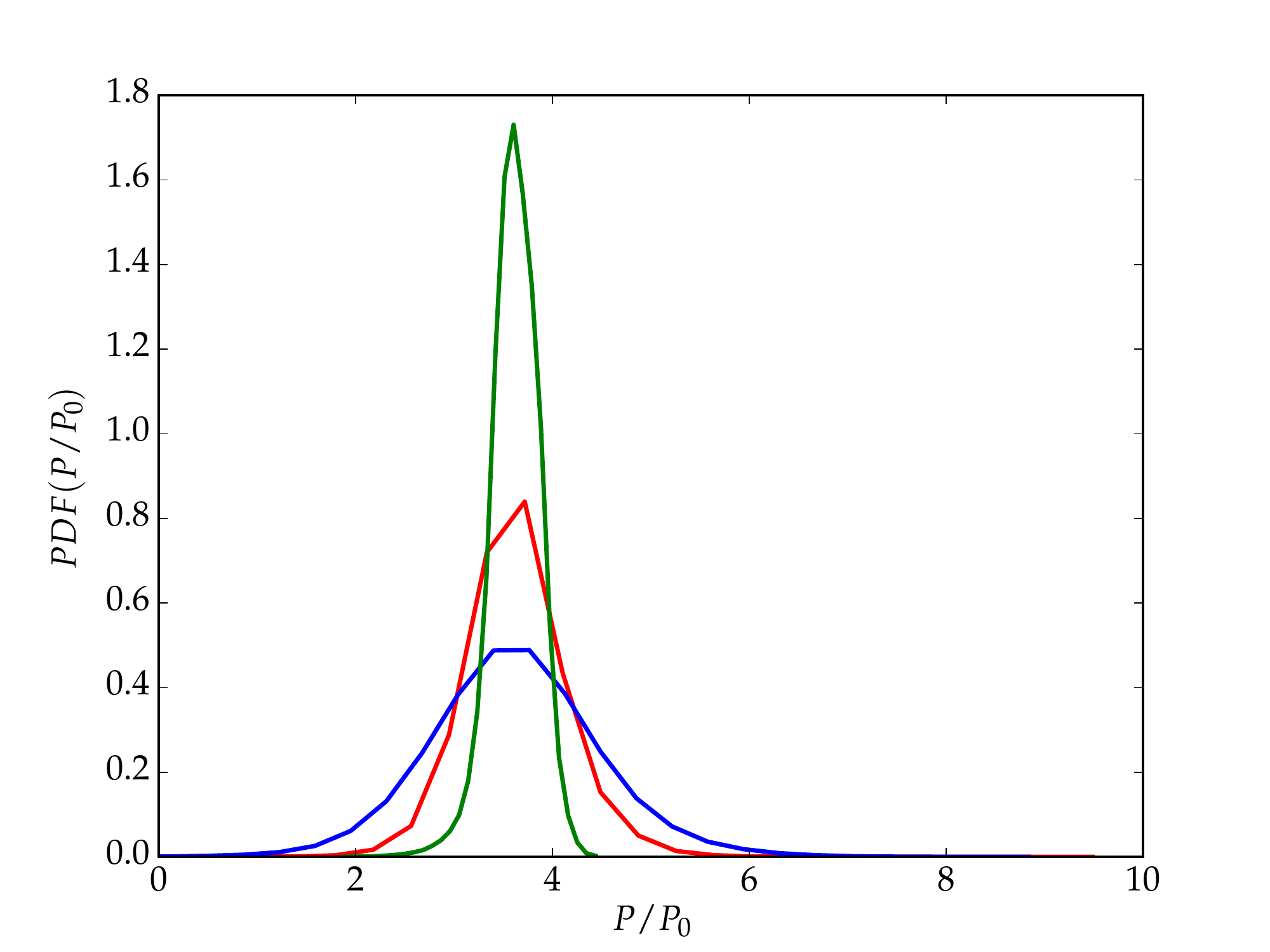}
\caption{The pressure probability distribution function from a
  fragmenting run at $t=40t_{\text{orb}}$. The green, red and green curves show the distribution at $t=0,4,14t_{\text{orb}}$.}
\label{fig::UN_pdf_P_f}
\end{figure}

\begin{figure*}
\includegraphics[width=8cm]{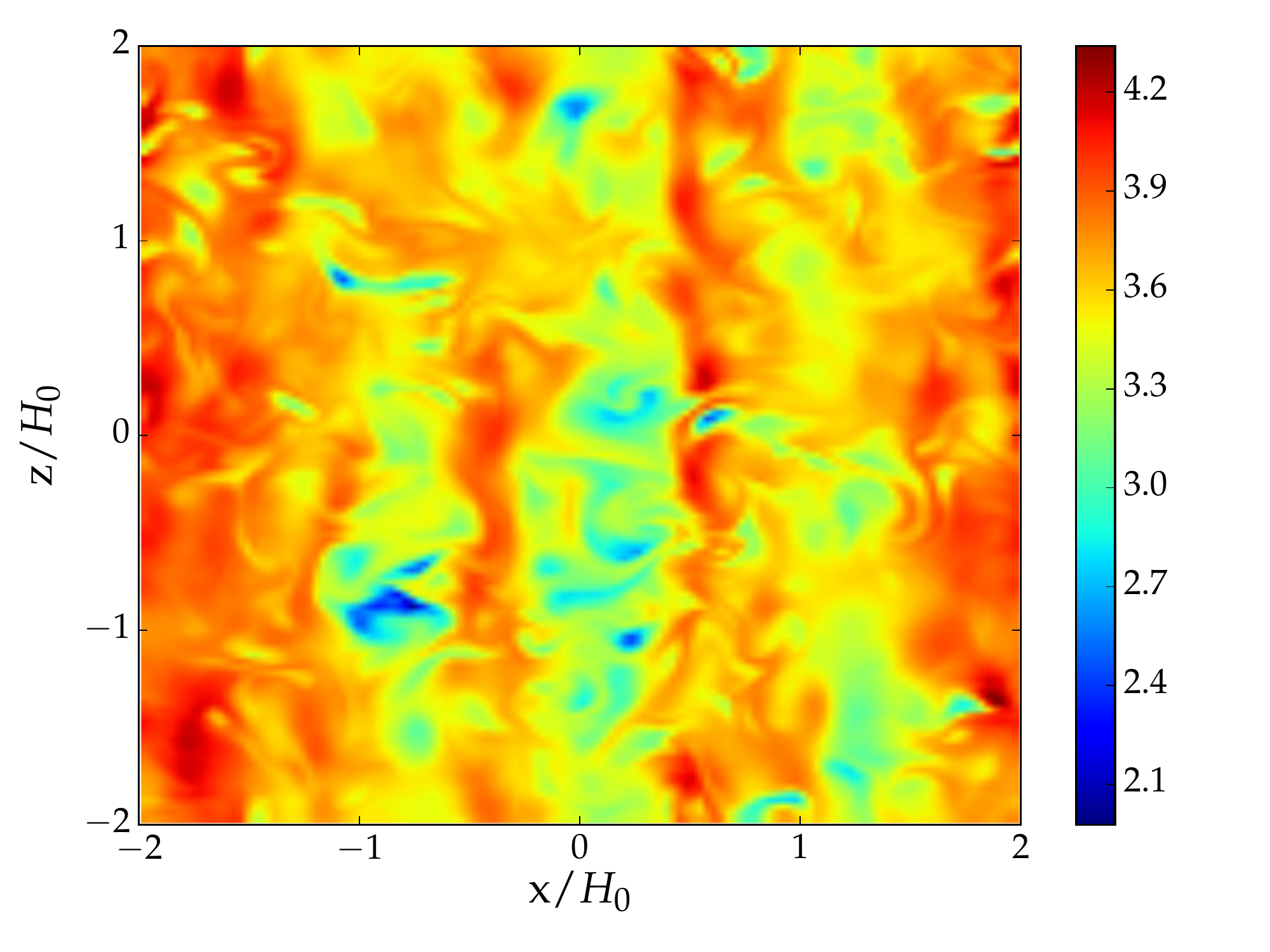}
\includegraphics[width=8cm]{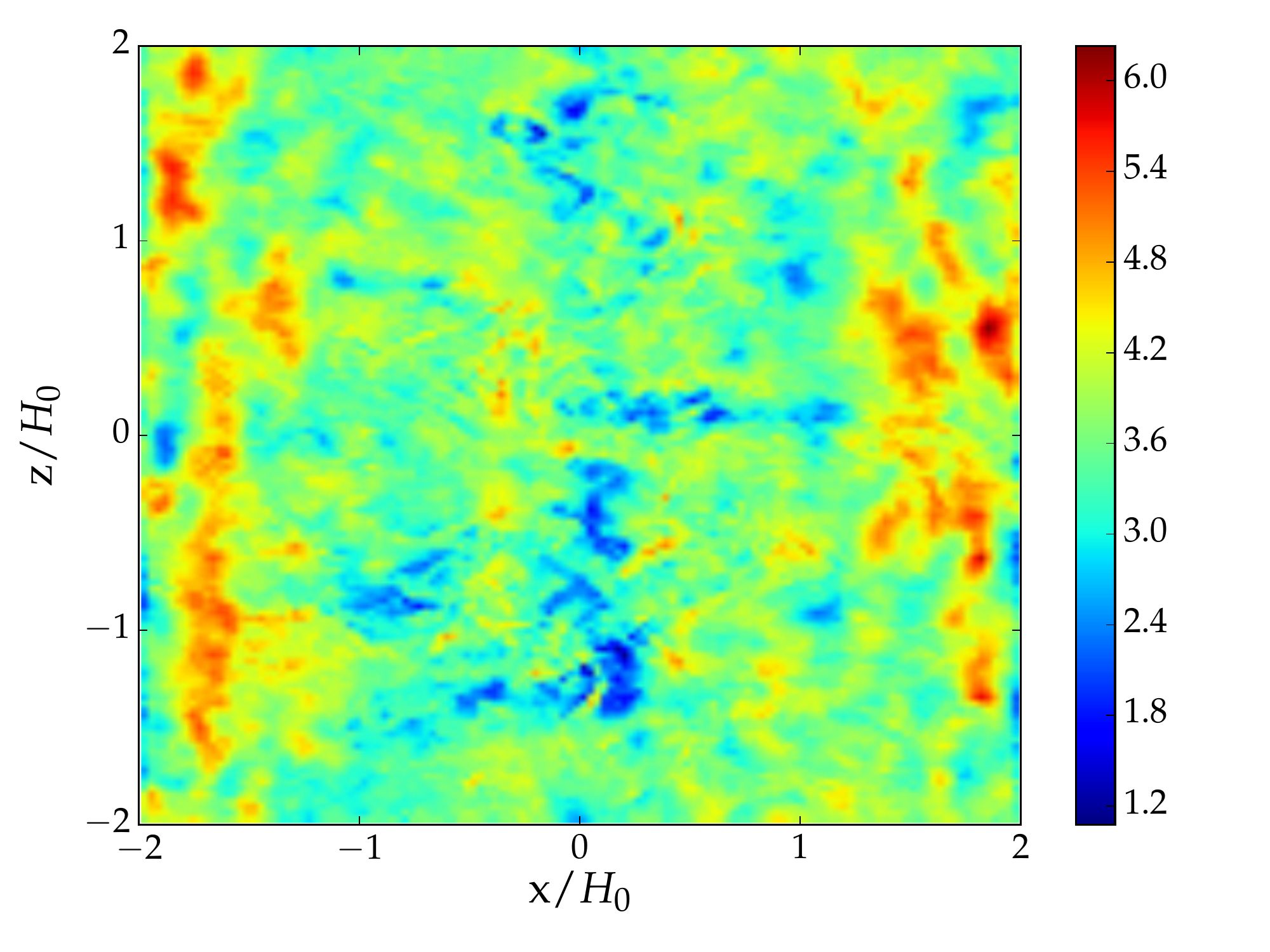}
\includegraphics[width=8cm]{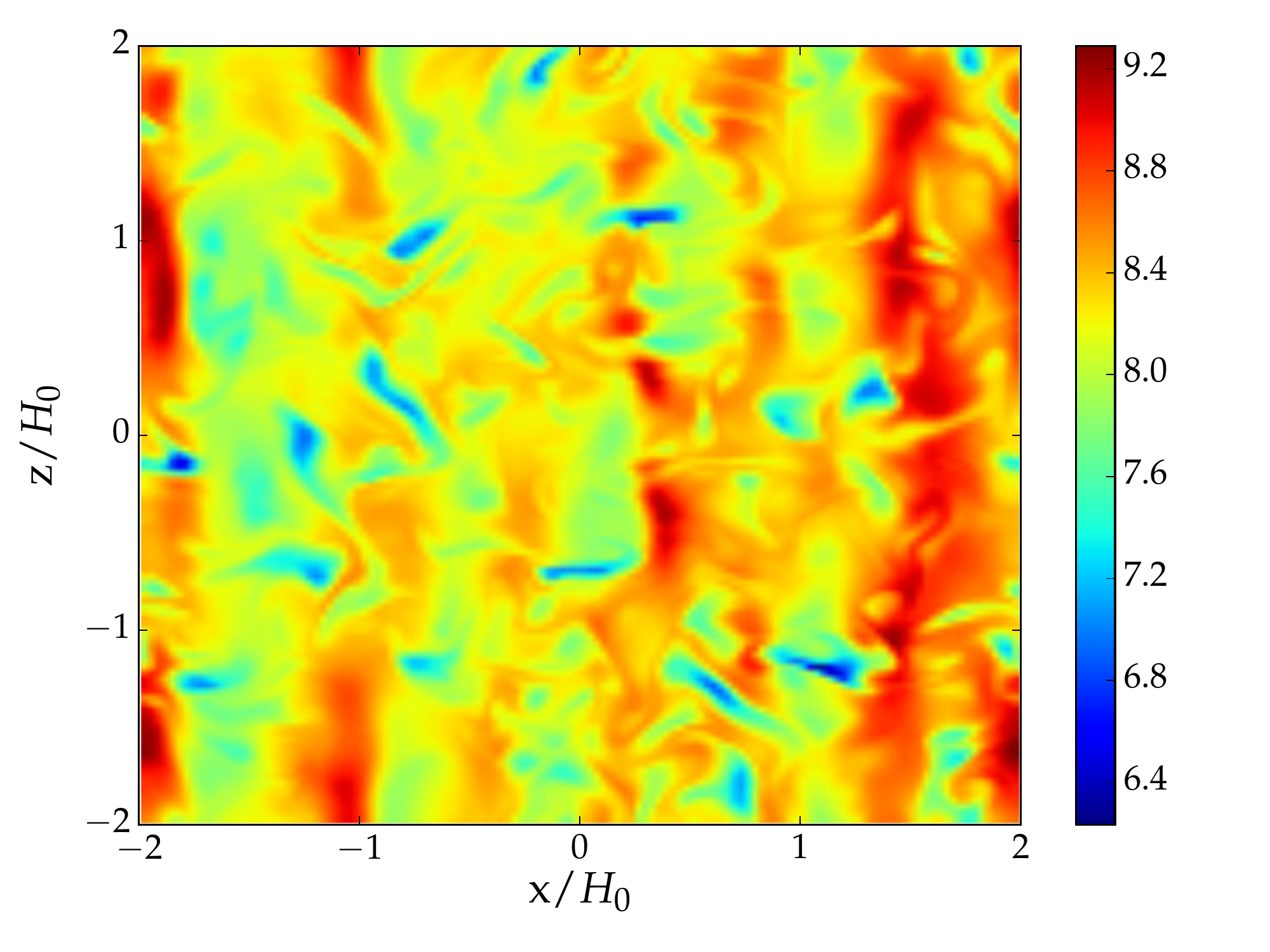}
\caption{Heat maps of $P/P_{0}$ in common $x-z$ slices. 
The top left panel is from the initial turbulent state, the top right
is from the 
the fragmenting simulation R3 at $t=4t_{\text{orb}}$. In the latter,
the difference in pressure between the coolest and hottest blobs is a factor 6.
For comparison we show a R2 simulation undergoing a slow heating
runaway at $t=340t_{\text{orb}}$ in the bottom panel. Here the
pressure difference is merely a factor 1.5. }
\label{fig::UN_snaP}
\end{figure*}

To show localised runaway and fragmentation, we consider a case
which we expect to be very unstable, choosing $m=0.1$ and
$\theta\approx1.84\times10^{-10}$ (simulation R3). In Fig.
\ref{fig::UN_pdf_P_f} we plot $PDF(P)$ at $3$ instances of time. 
The width of the distribution quickly increases,
indicating localised thermal runaway: initially Var($P/\langle
P\rangle$)$=0.0045$, but after $4$ and $14$ orbits this grows to
$0.019$ and $0.057$ respectively. During this time $\langle
P\rangle$ itself varies little, which 
emphasises that the box-averaged properties no longer give a
satisfactory description of the state of the system. Soon after the
final snapshot very small pressures occur resulting in the termination
of the simulation, preventing further exploration. In Fig.
\ref{fig::UN_snaP} we show $x$-$z$ slices in pressure before and during
thermal fragmentation. Prior to thermal fragmentation there are strong
acoustic waves propagating in the 
 radial direction. These appear to break up into
 patches that undergo rapid thermal runaway independent of each other.
Because this behaviour appears to be significantly nonlinear and
disordered, we do not attribute it to the action of a linear
instability mode with non-zero $k_x$ and $k_z$.

Though it is straightforward enough to achieve fragmentation in 
an unstratified local box, how likely is this in the 
inner regions of an X-ray binary? 
In this context the two major contributors to
mixing are radiative diffusion and turbulent advection. We first consider
radiative diffusion. In the hot dense gas, the opacity is largely
dominated by Thomson scattering with
opacity $\kappa_{T}=0.33\text{cm}^{2}\text{g}^{-1}$. The radiative diffusion
time across a length of $l$ is then $t_{\text{rad}}\sim
l^{2}/c\lambda$ where $c$ is the speed of light and $\lambda$ is the
mean free path. The instability timescale we estimate to be of order,
but bounded below, by the thermal timescale, so that
$ t_{\text{inst}} \gtrsim t_{\text{th}} \sim (\alpha\Omega)^{-1}$. 
If $t_{\text{rad}}\approx t_{th}$, we have the following condition on $l$
\begin{align}
\frac{l}{H}&\sim \left(\alpha\Sigma \kappa
  \frac{c_s}{c}\right)^{-1/2}, \\
&\sim
0.1\,\left(\frac{\alpha}{0.1}\right)^{-1/2}\left(\frac{\Sigma}{10^5\,
    \text{g cm}^{-2}}\right)^{-1/2}\left(\frac{T}{10^7 \,\text{K}}\right)^{-1/4},
\end{align}
where $\Sigma$ is surface density.
Regions separated by more than $l$ will be unable to mix
sufficiently well on the instability timescale. Typical values for an X-ray binary indicate that regions 
$0.1 H$ apart my in fact thermally fragment.

What about turbulent mixing?
We assume that the turbulent transport of heat by the MRI is similar
in efficiency to its transport of angular momentum (though this is a
point that has not been studied in detail). If we are permitted this
assumption then the turbulent diffusion time scale is
$t_{\text{turb}}\sim l^{2}/(\alpha  c_{s}H)$. If we next assume that the relevant eddies are of size $H$ 
(plausible if some form of MHD convection is operative)
then $t_{\text{turb}}\sim t_{\text{inst}}$ on these outer scales. These very rough
scalings indicate that turbulent heat transport is somewhat more
efficient than radiative transport, and moreover that it may be
sufficient to preclude fragmentation --- though a simple order of
magnitude treatment is unable to determine precisely when this might
occur. Only realistic simulations themselves can decide on this issue,
and in fact Jiang et al.~(2013) do not find fragmentation. Our
simulations
are marginally susceptible, in agreement with the above argument, but
they omit important physical effects such as buoyancy, which
may be crucial here, and enhanced compressibility effects in
radiation-dominated flow.

\section{Reduced stochastic models}

Because it is impossible to run a sufficient number of simulations to build
reliable statistics, especially regarding the distribution of $t_{\text{esc}}$,
we turn to simpler approximate models that illustrate
more fully the effects of the fluctuations on stability and which also
 permit analytical results.

We work primarily with the averaged energy equation \eqref{eqn::11}
but model the
fluctuating turbulent stress via a random function $\zeta$. Our model
is related to the logistic equation, and may be written as
\begin{equation}
\frac{dx}{d\tau}=\left[1+\zeta(\tau)\right]x^{q}-\Theta x^{m}.
\label{eqn::M2}
\end{equation}
This can be derived from \eqref{eqn::11} by an appropriate rescaling,
with $x$ and $\tau$ representing pressure/temperature and time,
respectively. Constant parameters are $q$, $m$, and $\Theta$.
To simplify the analysis while not losing much generality, we set 
$q=\Theta=1$ in much of what follows.

Equation \eqref{eqn::M2} admits the trivial steady state $x=0$ and the
more interesting equilibrium $x_{eq}=1$. In the
`laminar' case of $\zeta=0$ this equilibrium is unstable when $m<1$ 
with modes possessing the growth rate $1-m$.

For $\zeta(\tau)$ a random but \emph{continuous} function,
\eqref{eqn::M2} is the Bernoulli equation with analytic solution
\begin{dmath}\label{eq::sto}
\frac{x}{x_{eq}}=\phi(\tau)\left[\left(\frac{x_{0}}{x_{eq}}\right)^{1-m}-(1-m)\int^{\tau}_{0}\phi(s)^{m-1}ds\right]^{1/(1-m)}
\end{dmath}
where 
\begin{equation}
\phi(\tau)=\exp\left\{\tau+\int^{\tau}_{0}\zeta(s) ds\right\},
\end{equation} 
and $x_0$ is the initial value of $x$. As it stands, the analytic
solution is too unwieldy to be useful, even for basic prescriptions
for $\zeta$, but it does illustrate clearly the competition between
instability and stochasticity. These manifest as the two terms in
the exponent of $\phi(\tau)$. The first describes the deterministic 
exponential runaway, while the second stochastic term
potentially impedes this
tendency. In fact, if $\int^{t}_{0}\zeta(s)
ds$ behaves like a random walk,
then its standard deviation will be proportional to $\sqrt{\tau}$,
 and so on short to intermediate times the second stochastic term can
 outcompete the first instability term. On longer times, however,
 $\tau$ will always defeat $\sqrt{\tau}$ and the system will
approach the unstable laminar solution, given by
\begin{equation}\label{eq::det}
\frac{x}{x_{eq}}=\Bigg\{\left[\left(\frac{x_{0}}{x_{eq}}\right)^{1-m}-1\right]e^{(1-m)\tau}+1\Bigg\}^{1/(1-m)}.
\end{equation}

\subsection{Geometric Brownian motion (GBM)}

Before presenting an analysis of the full equation \eqref{eqn::M2},
it is worthwhile examining the simpler case of $\Theta=0$ and
$q=1$. The resulting system isolates cleanly all the main characteristics of
more realistic systems --- an unstable fixed point ($x=0$ now) and
stochastic noise --- while being analytically tractable. 
 Equation \eqref{eqn::M2} becomes
\begin{equation}
\frac{dx}{d\tau}=\left[1+\zeta(\tau)\right]x
\label{Eqn::M1}
\end{equation}
with initial condition $x_{0}>0$. For smooth random $\zeta$, the solution is
\begin{equation}
x(\tau)=x_{0}\phi(\tau).
\label{Eqn::GBM_gen}
\end{equation}

When studying stochastic dynamical systems, white noise is a convenient
choice for modelling the variability. For white noise to be a good
approximation, the fluctuation timescales of the system should be much
less than the characteristic time of interest. For MRI-driven
turbulence, this choice is not ideal given that the spectrum of
$\zeta$ has preferential frequencies. However, we use this as our
starting point as it makes a number of results especially clear. With this choice of
$\zeta$, Equation \eqref{Eqn::M1} must be written in differential form
\begin{equation}
dx=xd\tau+\sigma x dW
\label{Eqn::M1a}
\end{equation}
where $\sigma$ is the volatility coefficient (or noise amplitude) 
and $dW$ is white
noise. Here, for simplicity, we have interpreted the calculus in the Ito
sense.
 Equation \eqref{Eqn::M1a} actually describes geometric Brownian motion 
and is frequently used in financial modelling. Its solution is 
\begin{equation}
x(\tau)=x_{0}\exp\left\{\left[1-\frac{\sigma^{2}}{2}\right]\tau+\sigma W(\tau)\right\}.
\label{Eqn::M1sol}
\end{equation}

\begin{figure}
\includegraphics[width=9cm]{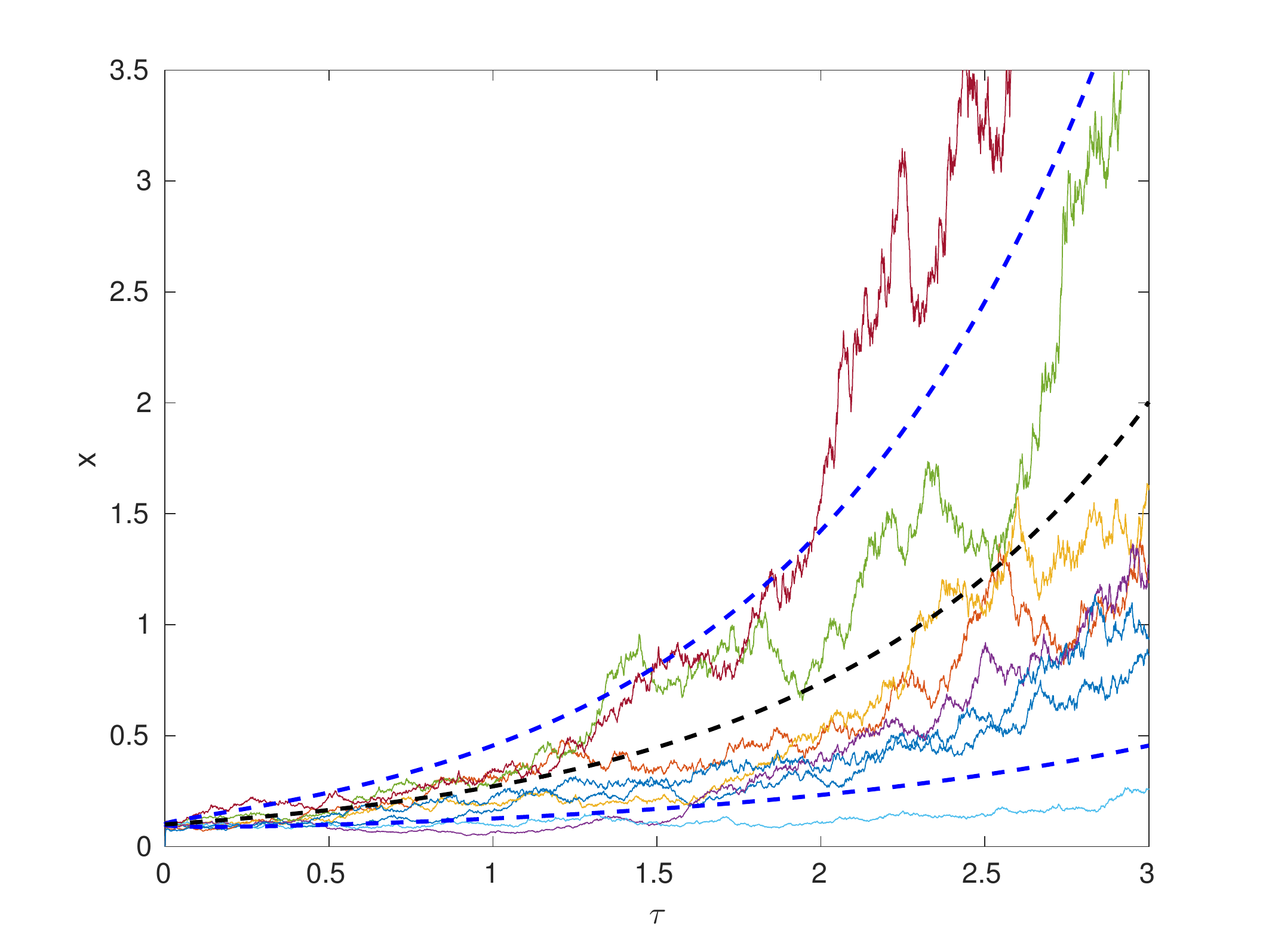}
\caption{ The solid curves are sample trajectories of geometric Brownian motion with $x_{0}=0.1$ and $\sigma=0.5$. The dashed black line is the laminar path and the blue dashed curves are the $10$th and $90$th percentiles.}
\label{Fig::GBM_Traj}
\end{figure}

Given that $x_{0}>0$, the solution remains strictly positive as a
result of the multiplicative form of the noise. We plot sample
trajectories in Fig. \ref{Fig::GBM_Traj} along with the $10$th and
$90$th-percentiles. When
calculating these trajectories we use the Euler-Maruyama method
(Kloeden \& Planten 1992). 
A feature of this collection of sample paths is
the wide variation between them, an attribute that is shared with the
simulations shown in Fig.
\ref{fig::UN_P_S}. They are also non-monotonic; fluctuations
`kick' the system towards or away from the fixed point. 
The light blue curve
 is particularly striking, exhibiting a trajectory that remain close
 to equilibrium,
$x(\tau)<0.25$, up to time $\tau\approx 3$. Note that at $\tau=3$ a
purely deterministic model would have predicted $x$ to be $\approx 2$,
an order of magnitude greater. 
The stochastic term in Equation \eqref{Eqn::M1sol} has
`balanced out' the deterministic drift, at least on these shorter
times.  

The reader may note that when $\sigma>\sqrt{2}$, the stability of
$x=0$ switches. It becomes an attractor, and the
system is stabilised. We stress, however, this effect is an artefact of
multiplicative white
noise in combination with the Ito calculus, and is not to be expected
in real turbulent systems. For instance, the stabilisation vanishes
in the Stratonovich calculus and/or with noise models with memory and
which are not multiplicative. We
certainly do not expect MHD turbulence to exhibit the combination of special
features that leads to this stabilisation. 
Indeed, it makes little physical sense that `shaking' an unstable
system more vigorously ultimately leads to zero
fluctuations. Moreover, 
the required amplitude of the fluctuations
must be extremely large, in our case this would require negative $\alpha$ which is impossible.
It is worth pointing out that the auto-regressive stochastic model
employed by Janiuk \& Misra (2012) shares the same stabilising property as white
noise in the Ito calculus; consequently, we view their
stabilisation of thermal instability as an artefact of their
model, and
not representative of a real fluctuating disk system.

The probability distribution of the solution trajectories may be obtained by
solving the associated Fokker-Planck equation. If $f(x,\tau)dx$ is the
probability of finding a path lying between $x$ and $x+dx$ at time
$\tau$, then 
\begin{align}
f(\tau,x)&=\frac{1}{\sigma x\sqrt{2\pi \tau}}\exp\left\{-\frac{(\log(x/x_{0})-\tau)^{2}}{2\sigma^{2}\tau}\right\}.
\end{align}
The expectation, $\mathbb{E}$, and variance, Var, can also be calculated 
\begin{align}
\mathbb{E}\left(x\right)&=x_{0}e^{\tau}, \\
\textrm{Var}\left(x\right)&=x_{0}^{2}e^{2\tau}\left(e^{\sigma^{2}\tau}-1\right).
\end{align}
The mean is independent of the volatility and
hence agrees with the laminar model. However, the variance contains a
$(e^{\sigma^{2}\tau}-1)$ factor which grows exponentially! This
means that as time progresses the expectation value grows less and less
meaningful because the distribution becomes increasingly wide and flat.
Overall, trajectories move away from the unstable equilibrium point,
but
individual trajectories can deviate from the laminar model dramatically.

Next we turn to the statistics of the escape time (or `last hitting
time'), which
 can be defined as follows:
\begin{equation}
t_{\text{esc}}=\max\left\lbrace \tau\geq0:x(\tau)= a\right\rbrace,
\end{equation}
where $a>x_{eq}=0$. This gives us the last time that a sample path is within the
interval $[0,a]$. Thus
$t_{\text{esc}}$ provides a measure for the effective instability timescale,
more accurate than the inverse of the laminar growth
rate. If the system was laminar ($\zeta=0$), however, the escape time
would be simply $t_{\text{esc}}^{\text{lam}}= \log(a/x_0)$. 

\begin{figure}
\includegraphics[width=9cm]{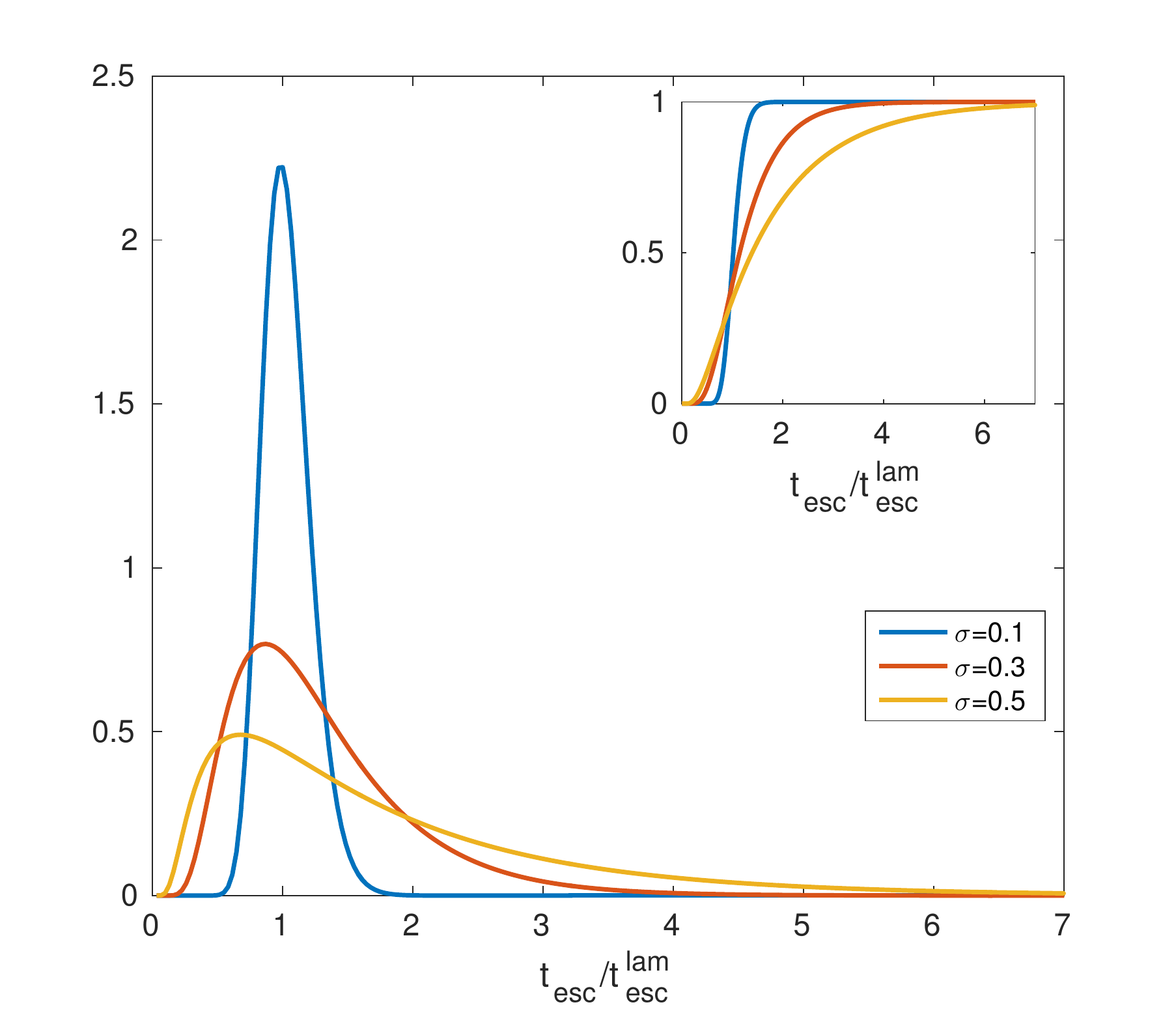}
\caption{Probability distribution function of the escape times for
  Geometric Brownian Motion with $x_{0}=0.1$ and $a=1.5$.}
\label{Fig::GBM_ET}
\end{figure}

For geometric Brownian motion, it is possible to
derive the probability distribution of $t_{\text{esc}}$ analytically (Kennedy 2010, Profeta 2010).
Because this is not a standard calculation we go through its details
in the Appendix. Denoted by $g(\xi)$ where $\xi=t_{\text{esc}}/t_{\text{esc}}^{\text{lam}}$, it is
a modified form of the Rayleigh distribution:
\begin{align}\label{escGBM}
g(\xi)&=\frac{1}{\sqrt{2\pi d^2 \xi}}  
\exp\left\{ -\frac{1}{2d^2}\left(\xi^{1/2}-\xi^{-1/2}\right)^2 \right\},
\end{align}
where the parameters $\sigma$, $x_0$, and $a$ have conveniently combined
into $d=\sigma\log(a/x_0)^{-1/2}$. In Fig. \ref{Fig::GBM_ET}
we plot $g(\xi)$ for a range of noise amplitudes.

The mean and variance of $g$ can be calculated analytically
\begin{align}
\mathbb{E}(\xi)&=1+d^2 \\
\textrm{Var}(\xi)&= d^2(1+ 2d^2).
\end{align}
It is clear that the mean of the distribution does not depart greatly from
the laminar escape time. Though the variance does increase with the
volatality of the noise, the spread is not especially
dramatic. Perhaps a more illuminating quantity is the kurtosis:
\begin{equation}
\text{Kurt}(\xi) = 3\,\frac{20d^2+9d^2+1}{(2d^2+1)^2},
\end{equation}
which varies from 3, when $d=0$, to 9, when $d=1$. A Gaussian
possesses a kurtosis of 3, so our escape time distribution can in fact
be exceptionally `fat-tailed', or `leptokurtic', meaning it generates
a significant number of outliers. Fig \ref{Fig::GBM_ET}
illustrates this point, with larger $\sigma$ giving rise to very
skewed and broad distributions. Inserted in the upper right
of the figure we
also plot the cumulative distribution function for various values of
$d$. For $d=1$, in particular, we see that only 30\% of systems
possess an escape time equal to or less than the laminar escape time ($\xi=1$),
while over 10\% of systems possess an escape time of four times or
more the laminar escape time. In fact, the probability that a
trajectory possesses a $t_{\text{esc}}/ t_{\text{esc}}^{\text{lam}}$ 
greater than some value
$\xi_0 $ may be approximated by the expression
$$ \mathbb{P}(\xi>\xi_0)\approx  \frac{d}{\sqrt{2\pi\xi_0}}
\text{exp}\left(\frac{2}{d^2}-\frac{\xi_0}{2d^2}\right),$$
for large $\xi_0$. When $d=1$ and $\xi_0=10$, the
probability is only about 1\%. However, this rises to 10\% when $d=2$,
indicating the strongly nonlinear dependence of the statistics on
noise amplitude.
In summary, through the statistics of the
escape time, one can observe stochasticity 
impeding instability over some initial period of the evolutio.

\subsection{Random logistic equation}

\begin{figure}
\includegraphics[width=9cm]{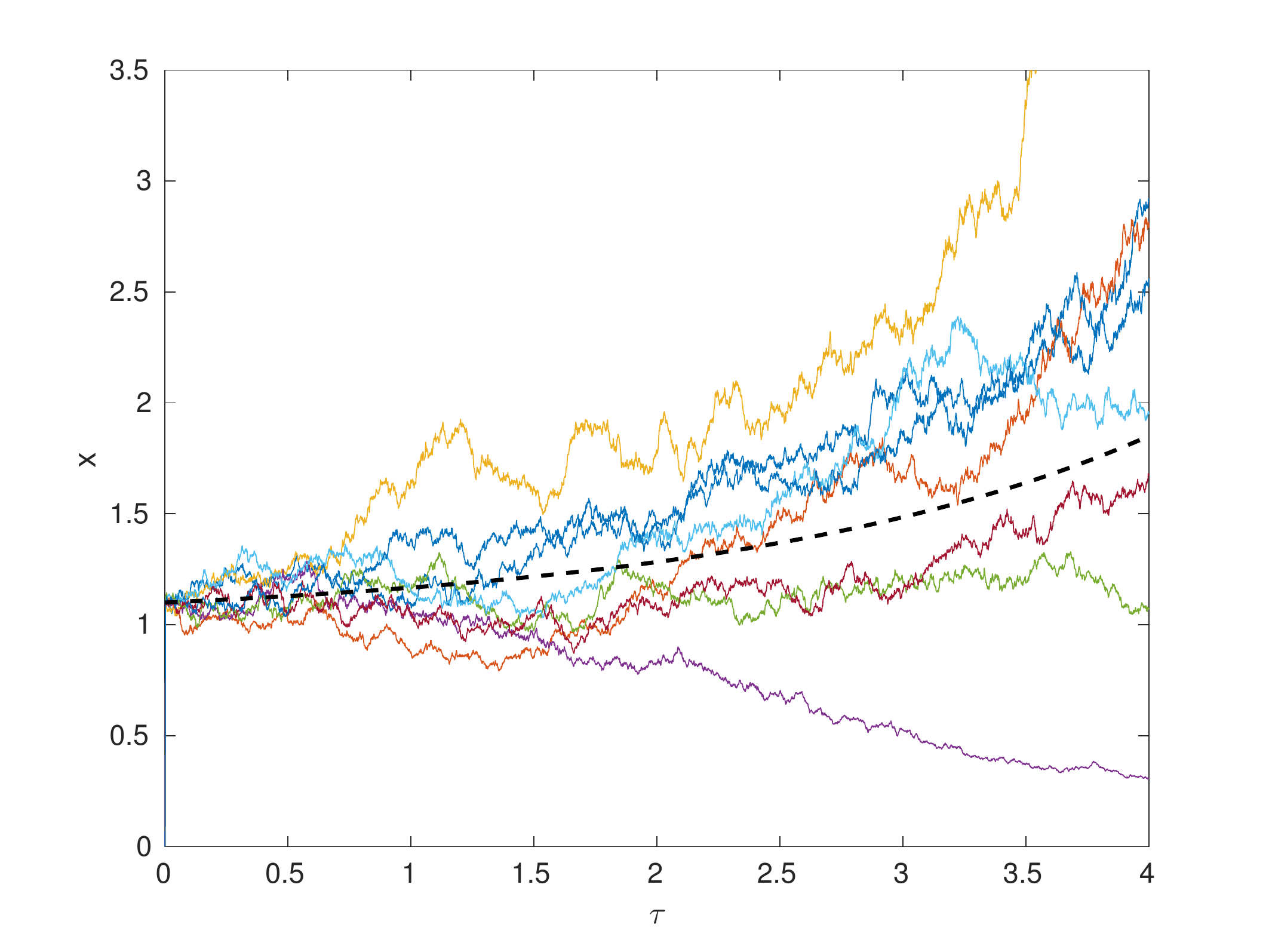}
\caption{Trajectories for the logistic model, $\sigma=0.2$ with $x_{0}=1.1$}
\label{Fig::TI_traj}
\end{figure}

We return to the more general problem by reintroducing the cooling
 term with $\Theta=1$
\begin{equation}
\frac{dx}{d\tau}=\left[1+\zeta(\tau)\right]x^q-x^{m}
\label{eqn::M2dfg}
\end{equation}
Again assuming white noise for $\zeta$ and setting $q=1$, 
we rewrite the ODE in differential form 
\begin{equation}
dx=\left(x- x^{m}\right)d\tau+\sigma x dW.
\end{equation}
The equation was solved numerically, and sample trajectories are shown in
Fig. \ref{Fig::TI_traj}. With a fixed initial condition, the system can either
undergo runaway heating or cooling. Note that the initial perturbation does
not give a good indication of which direction the sample paths
eventually go. The trajectories in
Fig. \ref{Fig::TI_traj} qualitatively resemble those in Fig.
\ref{fig::UN_P_S} from the MHD simulations. Both show trajectories
that remain close to their equilibrium values for multiple thermal
timescales along with trajectories that escape from the equilibrium
faster than the laminar model. One substantial difference exists
however: Fig. \ref{Fig::TI_traj} shows no indication that the
trajectories will be in general slower than the laminar model as  
is the case in Fig. \ref{fig::UN_P_S}. 

\begin{figure}
\includegraphics[width=9cm]{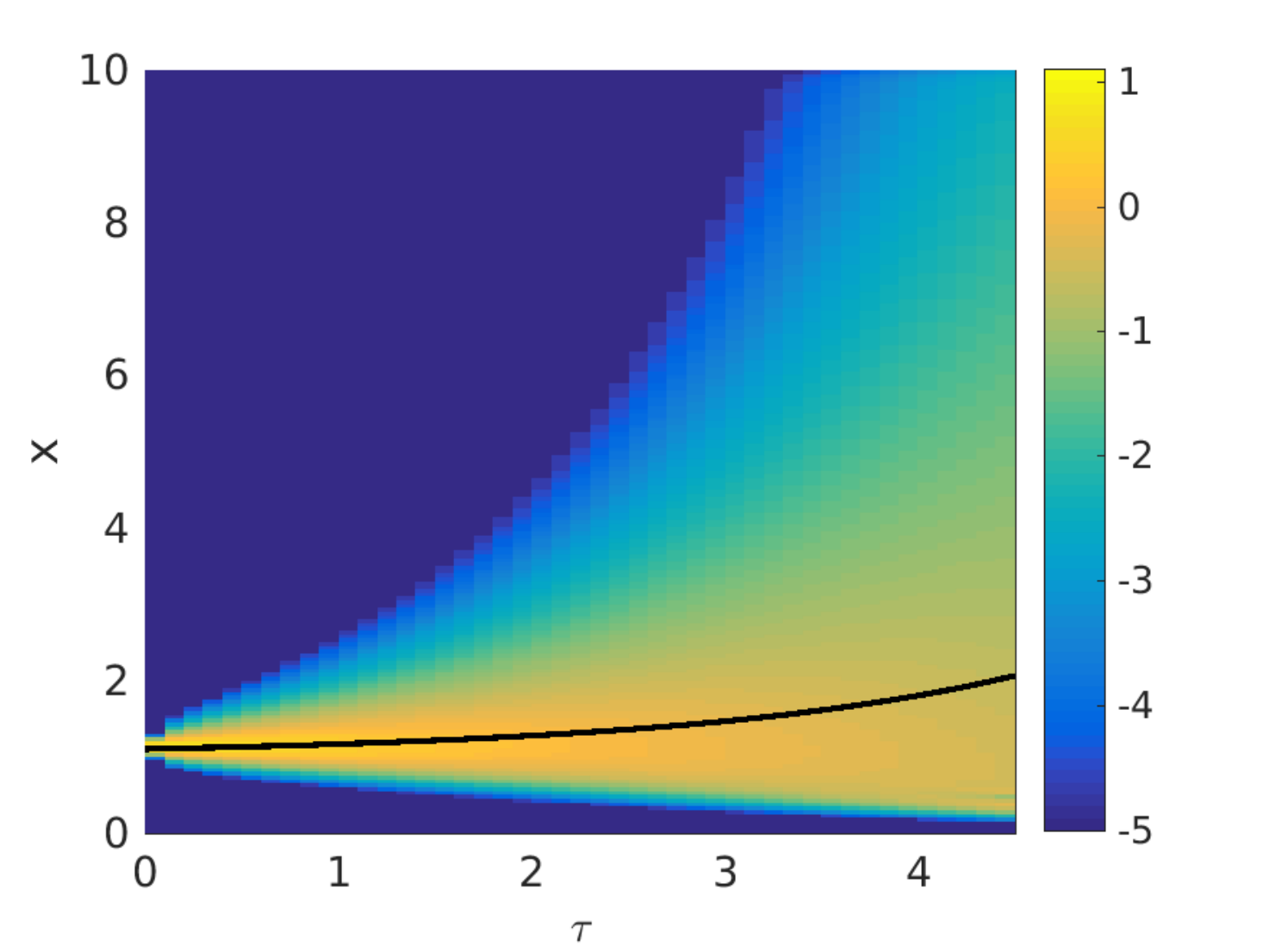}
\caption{Time evolution of the logarithm of the probability density
  function $f(\tau,x)$ for the logistic equation with $x_{0}=1.1$, $\sigma=0.1$.}
\label{Fig::TI_FP}
\end{figure}

Unlike geometric Brownian motion, it is difficult to analytically
calculate 
the probability 
 density function, and instead this is accomplished
by solving the Fokker-Planck equation numerically
\begin{equation}
\frac{\partial f(x,\tau)}{\partial \tau}=-\frac{\partial}{\partial
  x}\left[\left(x-x^{m}\right) f(x,\tau)\right]+\frac{\sigma^{2}}{2}
\frac{\partial^2}{\partial x^{2}}\left[x^{2}f(x,\tau)\right].
\end{equation}
In order to do this, an approximation for the initial distribution
must be made. Rather than using a Dirac-$\delta$ function as the initial
condition it is necessary to use a somewhat smoother function, 
a Gaussian distribution with variance $1000$. 
An example solution for the probability distribution $f$ is plotted
 in Fig. \ref{Fig::TI_FP}. As earlier, the distribution becomes
 increasingly wide and flat as time progresses, indicating the broad
 range in possible evolutionary paths.
 
\begin{figure}
\includegraphics[width=9cm]{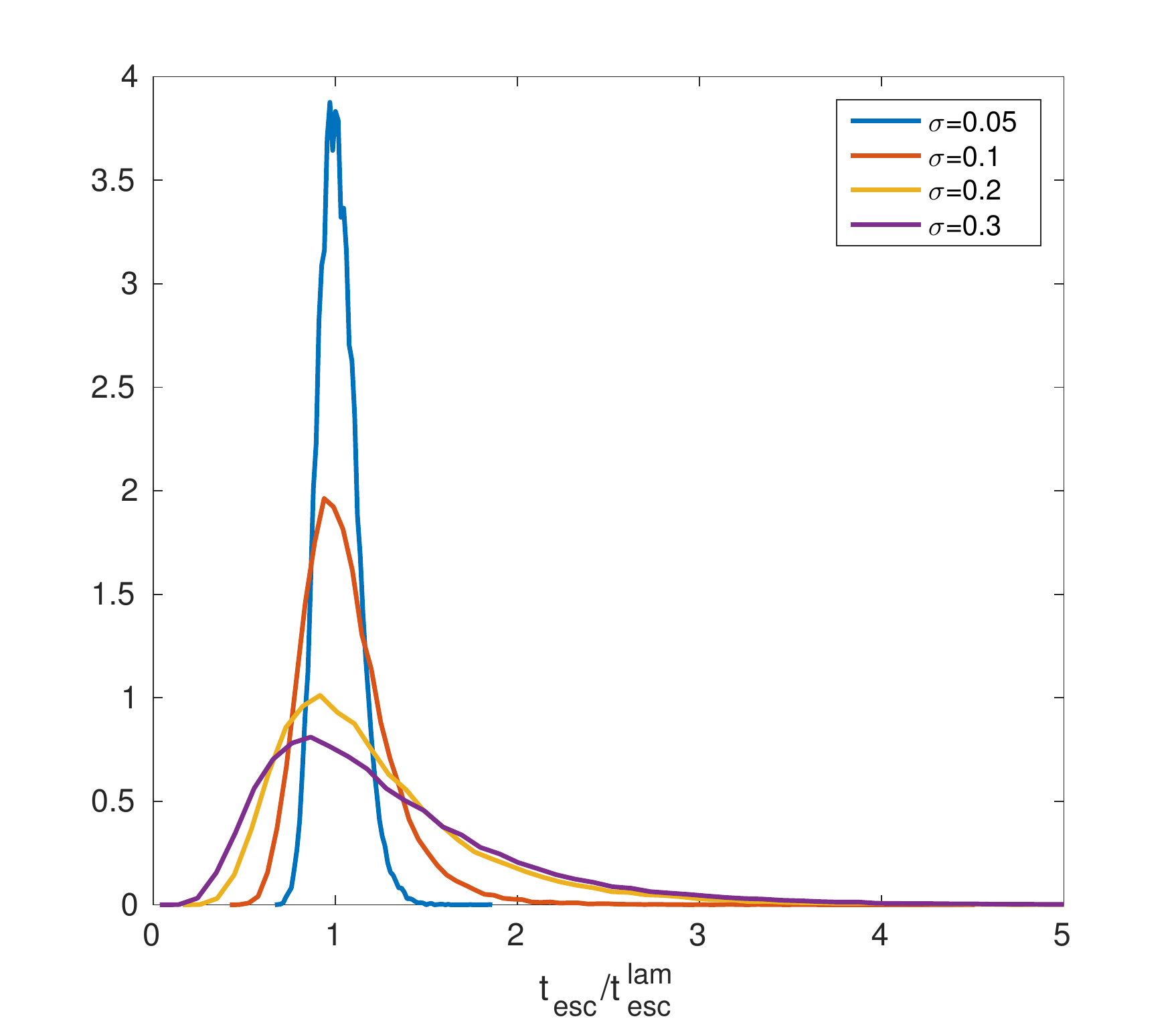}
\caption{Probability distribution of the escape time $g$
  for the logistic model, with $x_{0}=1.4$ with $a=4$ and $b=0.25$.}
\label{Fig::TI_ET}
\end{figure}

With the introduction of cooling comes the chance of a sample path
going to zero, hence, our escape time definition must be modified to include a lower boundary
\begin{equation}
t_{\text{esc}}=\sup\left\lbrace \tau\geq0:(x= a) \text{ or } (x=b)\right\rbrace
\end{equation}
where $a>x_{eq}>b$.
For a given sample path, the escape time is the last instance
when the solution is within the interval $[a,b]$, containing
$x_{eq}$. In Fig. \ref{Fig::TI_ET} we show estimated probability
density functions $g$ of $t_{\text{esc}}$ for a range of $\sigma$. These are
calculated from $100,000$ sample paths for each $\sigma$. As expected,
for small $\sigma$ we approach the laminar escape time, while as
$\sigma$ increases, $g$ undergoes asymmetric broadening,
shifting its maximum to
 lower $t_{\text{esc}}$ and developing a tail at long $t_{\text{esc}}$. Qualitatively
this behaviour mirrors that shown for
 geometric Brownian motion, Fig \ref{Fig::GBM_ET}. And so our
 conclusions for GBM carry over to the more realistic logistic case.
 
\subsubsection{Variable $q$ and $m$}

\begin{figure}
\includegraphics[width=8cm]{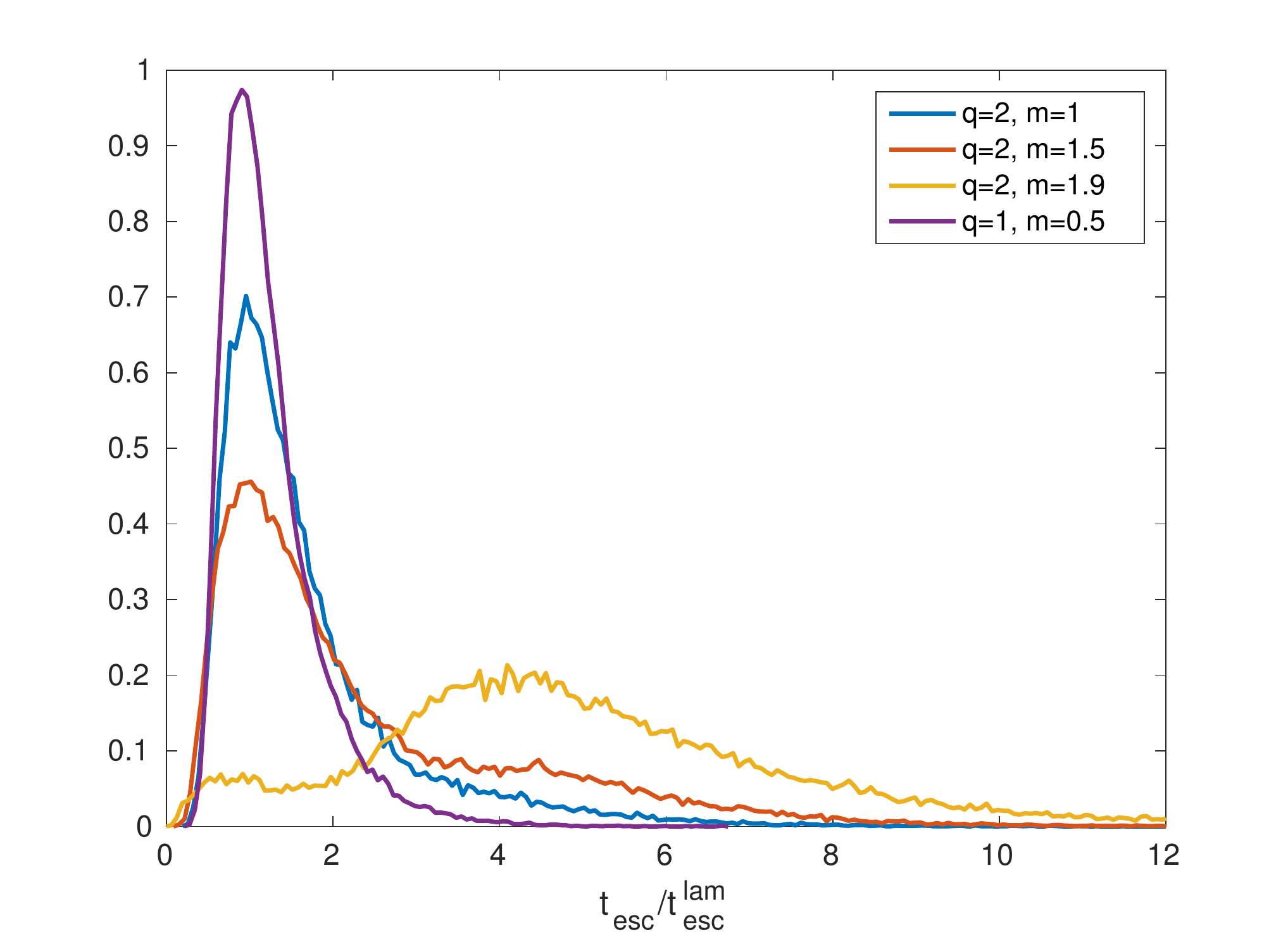}
\caption{Probability distribution of the escape time $g$
  for the logistic model with various $q$ and $m$ but keeping fixed: 
   $x_{0}=1.4$, $\sigma=0.2$, $a=4$, and $b=0.25$.}
\label{Fig::qm_ET}
\end{figure}

In Equation \eqref{eqn::M2}, $q$ and $m$ are free
  parameters which can be chosen to fit the system of interest. This
  flexibility
  allows it to approximately describe a range of different scenarios. 
  For
  instance, the classical Shakura \& Sunyaev disk (Shakura \& Sunyaev
  1973) can be represented with $q=2$ and $m=1$. Alternatively, $q$ and
  $m$ could be chosen so as to model the vertically stratified, radiation-pressure
  dominated simulations of Jiang et al. (2013, 2016). In this case a
  large surface density disk yields $q=1.6$ and $m=0.98$, while a less dense
  disk gives $q=1.9$ and $m=0.9$. Finally, when the temperature of the
  gas ensures the
  opacity is influenced by the `iron bump', the scaling of cooling with central
  pressure is found to greatly exceed that of the classical Shukura \&
  Sunyaev model and $m=1.89$ (Jiang et al.~2016). In
  Fig. \ref{Fig::qm_ET} we plot the probability distribution function
  of the escape time $g$ for various choices of $q$ and $m$. As $m$
  approaches $q$ from below, the escape time distribution becomes
  increasingly elongated to large
  $t_{\text{esc}}/t^{\text{lam}}_{\text{esc}}$. 
  The deterministic component
  weakens and the stochastic drift becomes more important, dominating the
  results.

\begin{figure}
\includegraphics[width=9cm]{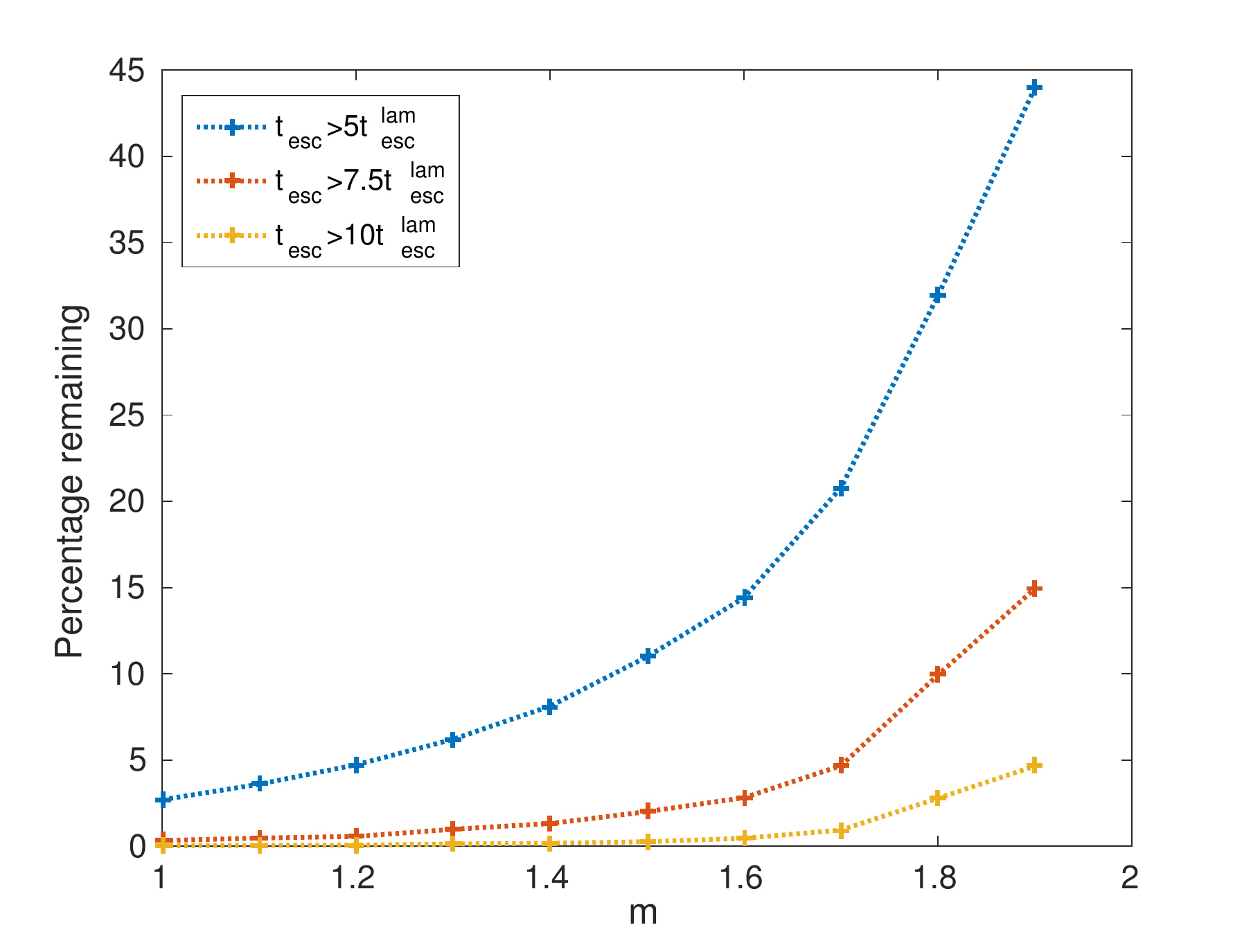}
\caption{The probability that $t_{\text{esc}}$ is longer than a
  multiple of $t_{\text{esc}}^{\text{lam}}$ as a function of $m$ for
  fixed $q=2$.}
\label{Fig::remain}
\end{figure}

 To better quantify this effect we compute the probability
  that a given trajectory possesses a $t_{\text{esc}}$ greater than
  various multiples of $t^{\text{lam}}_{\text{esc}}$ and plot these
  probabilities as a function of $m$ for fixed $q=2$. These results
  appear in Fig.~\ref{Fig::remain}. The curves show that as $m$
  approaches $q$ the escape time can be significantly enhanced. For
  $m=1.9$ there is approximately $45\%$ chance that it is five times
  the laminar prediction, and $5\%$ chance that is ten times greater.
  It should be noted that as $m$ approaches $q$ the laminar instability
  timescale itself can be significantly longer than the thermal 
  time $(\alpha\Omega)^{-1}$, which further separates the expected
  turbulent $t_{\text{esc}}$ from the thermal time. The stability
  uncovered in Jiang et al.~(2016), on the timescales of their
  simulations, can be easily explained by the enhanced delay witnessed
  in such
  marginally unstable systems.

\begin{figure}
\includegraphics[width=8cm]{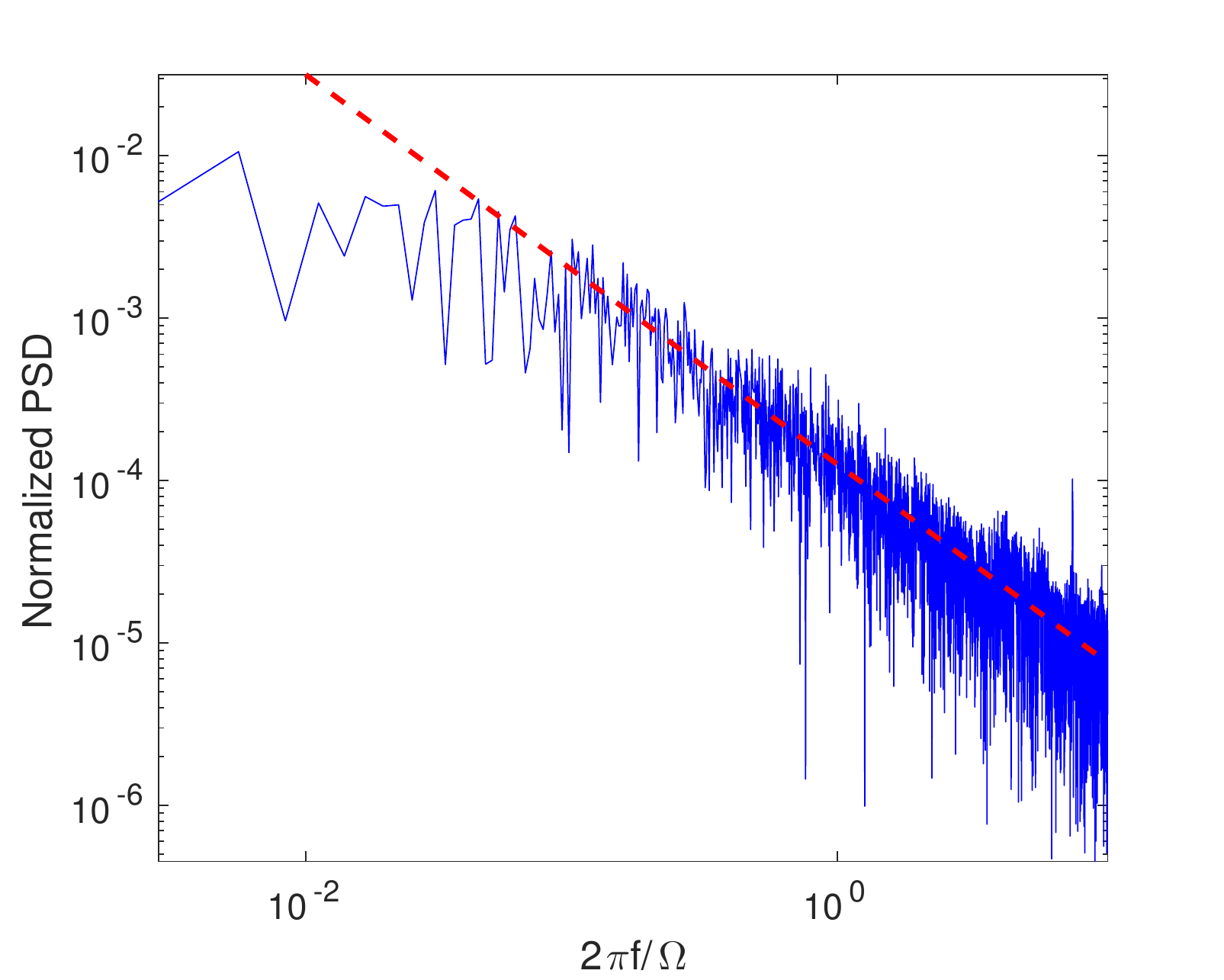}
\caption{Time averaged power spectral density normalised by $P_{0}$ from simulation R1. Slope $\propto f^{-1.2}$}
\label{Fig::PSD}
\end{figure}

\begin{figure}
\includegraphics[width=8cm]{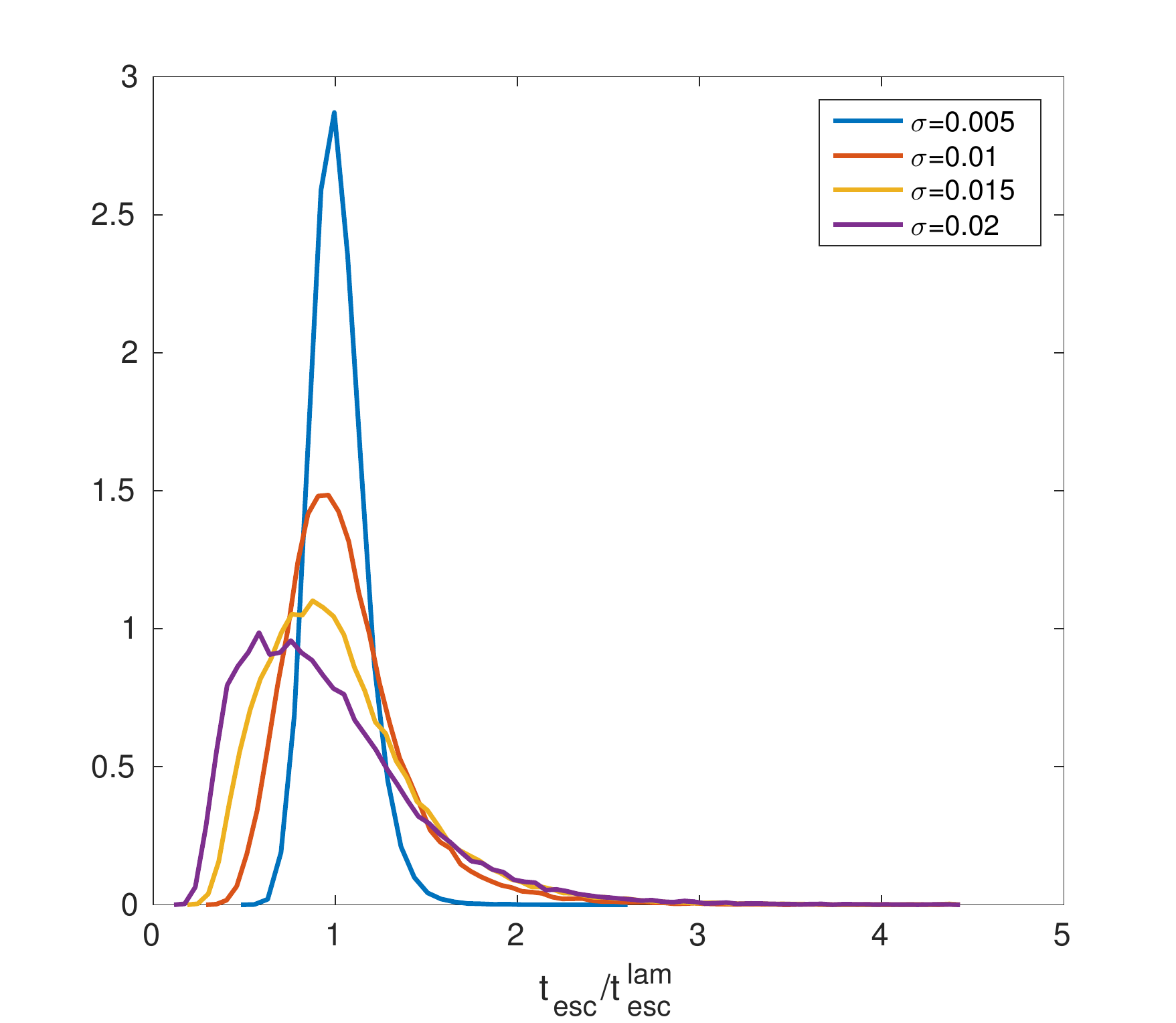}
\caption{Escape time distribution from our power spectral density model with $q=0.5$ and $m=0.25$}
\label{Fig::PSD_ET}
\end{figure}

\subsection{MHD power spectral density}

The next step is to replace the white noise with the power spectral
density (PSD) of $\Pi_{xy}$ calculated from the MHD simulations of
Section 4. 
In Fig. \ref{Fig::PSD} we show the PSD of $\Pi_{xy}$ from
simulation R1. We then represent
the fluctuating term in Equation \eqref{eqn::M2} by
 \begin{align}
\zeta(\tau)&=\frac{\sigma}{N}\Sigma_{i=1}^{200} \alpha_{i}(f_{i}) \cos\left( 2\pi f_{i}+\phi_{i}\right),\label{eqn::PDS} \\
N&=\left(\Sigma_{i=1}^{200}\alpha_{i}(f_{i})^{2}\right)^{1/2}
\end{align} 
where $f_{i}$ are the frequencies, logarithmically spaced between $\left[0.05,
 750\right]$, and $\phi_{i}$ are phase shifts, chosen randomly
from a 
uniform distribution on the interval $\left[0, 2\pi\right]$. 
Finally, the constant amplitudes $\alpha_{i}(f_{i})$ are
calculated from a two slope power law that fits our
MHD simulations (constant for $2\pi f/\Omega<0.1$ and $\propto
f^{-1.2}$ otherwise), which is then multiplied by a random number
generated from a uniform distribution on the interval $[0,1]$.

To improve the comparison with the simulations R2a-R2h, we set $q=0.5$
and $m=0.25$ 
and thus our model equation is
\begin{equation}
\frac{dx}{d\tau}=\left[1+\zeta(\tau)\right]x^{0.5}-x^{0.25},
\label{eqn::M2l}
\end{equation}
where $\zeta(\tau)$ is defined in Equation $\eqref{eqn::PDS}$. This
equation is evolved forward in time $50,000$ times for each choice of $\sigma$
in order to derive adequate statistics, especially for the escape time.

We show the resulting escape time distribution in 
Fig. \ref{Fig::PSD_ET}. We find similar behaviour to both
the GBM and the random logistic equation. As $\sigma$ increases, the
distribution broadens and its tail at large $t_{\text{esc}}$
increases. We estimate that $\sigma\sim0.015$ gives approximately the
correct fluctuation amplitude when compared to the stress in
Fig. \ref{Fig::r1_p}. For this choice,  Fig. \ref{Fig::PSD_ET} 
shows a clear tail reaching $t_{\text{esc}}/t_{\text{esc}}^{\text{lam}}\sim2.5-3$.

We
conclude that fat-tailed distributions are a generic feature of the
escape time in turbulent but thermally unstable systems. Such systems
produce a broad range of outcomes, and instability can be delayed for
several instability timescales. Being fat-tailed they also exhibit
significant outliers --- systems that `hang around' for surprisingly
long times before wandering away.

\section{Discussion and Conclusions} \label{Sec::TIDC}

We have performed a set of idealised shearing box simulations of the MRI
in order to explore the effects of
turbulent variability on thermal instability. Our main aim was to
check if turbulence interferes with the thermal runaway
predicted by laminar theory. In our
simulations heating comes from numerical dissipation while
a power law cooling imitates optically thin radiative
cooling. Relatively large computational domains are used in  
order to ensure a strong dependence of stress on pressure.  

Simulations with
an expected stable thermal equilibrium are found to fluctuate around
their fixed points. When the cooling power law exponent is decreased,
and the laminar analysis predicts instability, our simulations
indeed show thermal runaways. However, the
system trajectories deviate from the laminar
theory. The turbulent fluctuations can stall the onset of instability
in a large number of cases, for multiple thermal times. To better
account for the instability timescale, we introduce the concept of the
escape time which we define to be the last time the system leaves
an interval encapsulating the equilibrium in phase space.
Our simulations show a large range of escape
times ranging from $\sim 1/4$ to $\sim 5$ times the laminar thermal
timescale, for relatively narrow intervals.

Further reducing the cooling power law exponent results in disk
fragmentation. This is a due to localised imbalances between
heating and cooling; the instability timescale is shorter than the
mixing timescale and hence distant pockets of fluid evolve
independently. 
Very rough estimates indicate that thermal fragmentation on length scales larger
than $H$ are at best marginally possible in the inner regions 
of X-ray binaries.

To better understand our results we construct a probabilistic theory centred
on 
simple stochastic equations that approximate the box-averaged thermal
equation. These present us with a much larger sample of possible
system outcomes to analyse than the MHD simulations can afford.
First we analyse geometric Brownian motion (GBM),
which possesses the main ingredients of interest (an
unstable fixed point and stochastic variations), while remaining
analytically tractable. The distribution function of $t_{\text{esc}}$ exhibits
a variance proportional to the square of the noise amplitude and 
considerable kurtosis. In general the distribution is `fat-tailed',
permitting many instances of delayed thermal instability, and outliers
for whom the escape time can be $\sim 10$ thermal times.   
Our second model introduces a logistic
term to incorporate a power law cooling, and a third model
replaces the white noise with the power spectral density of
the stress from a thermally stable shearing box simulation. In both
cases we obtain qualitatively similar behaviour to before, 
which instils confidence that its behaviour is generic to
systems sustaining noise and instability. GBM
may be thought of as the
model equation for such systems.

In our GBM model, very large 
amplitude noise can stabilize the thermal instability, but we 
believe this is physically implausible. In fact, this behaviour
arises from the special combination of Ito calculus and multiplicative white
noise. It is not generic. 
While GBM is a convenient model for
 `unresolved' turbulence, this dynamical peculiarity must be discarded when
applying results to real systems. The stochastic process
employed by Janiuk \& Misra (2012) shares the same properties and thus
suffers
the same dynamical artefact. The stabilisation witnessed in their
simulations we hence view as unphysical.

The bulk of X-ray binary observations show no indication of limit
cycles that could correspond to radiation-pressure induced 
thermal instability
(Gierli\'{n}ski \& Done 2004). Notable exception are GRS 1915+105
(Done et al.~2004) and HLX-1 (Sun et al.~2016). We show that it is
unlikely that
turbulent fluctuations alone stabilise these disks. 
However, turbulence can delay and weaken instability.
This weakening might be
significant
in combination with additional
stabilising mechanisms, for example: energy lost in the disk corona or by outflows,
magnetic buoyancy effects, alteration of the pressure-stress
relationship by strong magnetic fields, or opacity shifts near the
iron bump (Svensson \& Zdziarski 1994, Sadowski 2016, Jiang and Stone
2016). Finally, separate and important physical effects may arise from the
global nature of the flow, especially from accretion. The viscous timescale
is probably longer than a typical $t_{\text{esc}}$, as measured in this
paper. But if further delayed by additional physics, $t_{\text{esc}}$ 
could in some circumstances 
approach the
accretion time. If this occurs then the classic limit cycle behaviour expected
from thermal instability could well be impeded, and/or pushed to smaller
radii.

\section*{Acknowledgements}
The authors thank the anonymous reviewer for
a set of helpful comments. They also thank Yanfei Jiang
and Jim Stone for discussions that helped sharpen the paper.
This work was partially funded by STFC
grants ST/L000636/1 and ST/K501906/1.
Some of the simulations were run on the
DiRAC Complexity system, operated by the University of Leicester
IT Services, which forms part of the STFC DiRAC HPC Facility
(www.dirac.ac.uk). This equipment is funded by BIS National E-
Infrastructure capital grant ST/K000373/1 and STFC DiRAC Op-
erations grant ST/K0003259/1. DiRAC is part of the UK National
E-Infrastructure.

\bibliographystyle{apalike}

\begin{appendix}

\section{Escape time distribution for GBM}\label{Ap::B}

Following Kennedy (2010), we derive the probability density function (PDF) for the escape
time, or last
hitting time, for geometric Brownian motion. To achieve this, we
calculate the PDF of the last hitting time of Brownian motion
with drift and then perform a change of variables to obtain the last
hitting time for GBM.  

In the appendix
$W_{\tau}$ will refer to a standard Brownian motion, 
the amplitude of the fluctuations can be represented by a pre-factor
$\sigma$,
which we refer to as the `volatility'.
To begin, consider the
\emph{first} hitting time of a random function $x(\tau)$,
denoted by $M_a$ and defined by
\begin{equation}
M_{a}=\inf\left\{\tau\geq 0:x(\tau)=a\right\}.
\end{equation}
(Here the infimum can be thought of as the minimum.) Physically this
represents the first time the function $x$ reaches the value
$a$. In the special case of $x$ corresponding to
 standard Brownian motion with drift, $x(t)=W_{\tau}+\epsilon \tau$,
the probability density function, $f_{M_a}$, of $M_a$ is
\begin{equation}
f_{M_{a}}(\xi)=\frac{a}{\sqrt{2\pi \xi^{3}}}\exp\left\{-\frac{1}{2}\left(\epsilon \sqrt{\xi}-a/\sqrt{\xi}\right)^{2}  \right\}.
\label{eqn:BMwDPDF}
\end{equation}
See Kennedy (2010) for a derivation. 
With this result, we can then determine the \emph{last} 
hitting time of Geometric Brownian motion with drift. 

Let $T_{-\mu \tau+b}$ denote the last time that the 
standard Brownian motion hits 
the line $(-\mu \tau +b \geq0)$, for some constant $b$. 
Then the probability density function of $T_{-\mu \tau+b}$ is given by 
\begin{equation}
f_{T_{-\mu \tau+b}}(\xi)=\frac{\mu}{\sqrt{2\pi \xi}}\exp\left\{-\frac{1}{2}\left(\mu\sqrt{\xi}-b/\sqrt{\xi}\right)^{2}\right\}.
\end{equation}
To see why this is true consider the probability
that $T_{-\mu \tau+b}> \xi$, for some $\xi$:
\begin{align*}
\mathbb{P}\left(T_{-\mu \tau+b}>\xi\right)&=\mathbb{P}
\left(W_{s}=b-\mu s,\text{ for some  }s>\xi\right) \\
&=\mathbb{P}\left(sW_{1/s}=b-\mu s,\text{ for some  } s>\xi\right) \\
&=\mathbb{P}\left(W_{u}=b u-\mu,\text{ for some  }u<\frac{1}{\xi}\right) \\
&=\mathbb{P}\left(M_{-\mu}<\frac{1}{\xi}\right).
\end{align*}
In the second equality we have used the fact that $\{sW_{1/s}\}$ is
also a
Brownian motion. The probability distribution function of 
$T_{-\mu \tau+b}$ is given by the $\xi$
 derivative of the above expression. By considering the probability
 distribution function of $M_{-\mu}$ where the drift is $-b$ and the
 threshold $-\mu$, we conclude that
\begin{equation}
f_{T_{-\mu \tau+b}}(\xi)=\frac{1}{\xi^2}f_{M_{-\mu}}\left(\frac{1}{\xi}\right)
\end{equation}
from which the result follows.
Notice that $f_{T_{-\mu \tau+b}}$ is identical to the 
probability density function of the last time Brownian motion 
with drift $\mu$ hits the line $b$.  

Let us now finally define the last hitting time, or escape time, for a
random function $x(\tau)$:
\begin{equation}
T_{a}=\sup\left\{\tau\geq0:x(\tau)=a\right\},
\label{Eqn::BMD_ET}
\end{equation}
for some threshold $a$.
If $x$ corresponds to Brownian motion with drift 
$x(\tau)=\nu\tau+W_\tau$, for constant $\nu>0$, then the probability
distribution of $T_a$ is
\begin{equation}\label{fbdrift}
f_{T_{a}}(\xi)=\frac{\nu}{\sqrt{2\pi \xi}}
\exp\left\{-\frac{1}{2}\left(\nu\sqrt{\xi}-a/\sqrt{\xi}\right)^{2}\right\}. 
\end{equation}
If however, $x(\tau)$ is a GBM
\begin{equation}
x(\tau)=x_{0}e^{\sigma W+\mu \tau},
\end{equation}
and $x_{0}$ a constant, then the distribution of $T_b$, for some
constant threshold $b$, is easy to
obtain from \eqref{fbdrift}. Consider the probability for GBM that
$T_a< \xi$, for some $\xi$. This corresponds to
\begin{align*}
\mathbb{P}(T_a\leq \xi)&=\mathbb{P}\left(x(\tau)\geq b, \forall \tau\geq \xi\right) \\
&= \mathbb{P}\left(\mu \tau +\sigma W_{\tau}\geq \log\left(\frac{b}{x_{0}}\right), \forall \tau\geq \xi\right) \\
&= \mathbb{P}\left(\frac{\mu}{\sigma} \tau +W_\tau\geq
  \frac{1}{\sigma}\log\left(\frac{b}{x_{0}}\right),
  \forall \tau\geq \xi\right)
\end{align*}
which is precisely the probability for Brownian motion with drift, but with
 $\nu=\frac{\mu}{\sigma}$ and $a=\frac{1}{\sigma}\log(b/x_{0})$. This
 supplies us with an expression for the distribution of $T_{a}$ for
 GBM. A rescaling of $\tau$ and a renormalisation obtains Equation \eqref{escGBM}.

\end{appendix}

\end{document}